\documentclass[aps,onecolumn,preprintnumbers,nofootinbib,superscriptaddress]{revtex4}
\usepackage{standalone}
\usepackage{tikz}
\usepackage{amsmath}
\usepackage{graphicx}
\usepackage{amsfonts}
\usepackage{array}
\usepackage{amsthm}
\usepackage{bm}
\usepackage{palatino}
\usepackage{mathpazo}
\usepackage{supertabular}
\usepackage{subfig}
\usepackage[breaklinks]{hyperref}
\usepackage{color}
\usepackage[labelfont=bf,labelsep=quad,justification=justified]{caption}
\usepackage{epstopdf}
\allowdisplaybreaks
\begin{document}

	\title{Bound state solutions with a linear combination of Yukawa plus
		four-parameter diatomic potentials using path integral approach: Thermodynamic properties}

	\author{ Mohamed Améziane Sadoun}
	\email{m.sadoun@univ-bouira.dz }
	\affiliation{\it Laboratory of Materials Physics and Optoelectronic Components, Department of Physics, Faculty of Exact Sciences, University of Bouira, 10000 Bouira, Algeria}
	
	\author{ Redouane Zamoum}
	\email{r.zamoum@univ-bouira.dz (corresponding- author)}
	
	\affiliation{\it Laboratory of Materials Physics and Optoelectronic Components, Department of Physics, Faculty of Exact Sciences, University of Bouira, 10000 Bouira, Algeria}
	
	\author{ Abdellah Touati}
	\email{touati.abph@gmail.com}
	
	\affiliation{\it Department of Physics, Faculty of Exact Sciences, University of Bouira, 10000 Bouira, Algeria}

	\date{\today}

	\begin{abstract}
		In this paper, we investigate the approximate analytical bound states with a linear combination of two diatomic molecule potentials, Yukawa and four parameters potentials, within the framework of the path integral formalism. With the help of an appropriate approximation to evaluate the centrifugal term, the energy spectrum and the normalized wave functions of the bound states are derived from the poles of Green’s function and its residues. The partition function and other thermodynamic properties were obtained using the compact form of the energy equation.
	\end{abstract}
	
	%\pacs{04.20.Fy, 04.20.Jb, 04.40.Nr, 04.70.Bw}
	
	\keywords{Green's function, Yukawa potential, Four-parameter diatomic potential, Energy spectrum, Wave functions, Thermodynamic properties}
	
	\maketitle
	
	%---------------------------------------------------------------------------------------%---------------------------------------------------------------------------------------

	%---------------------------------------------------------------------------------------
	\section{Introduction}
	%---------------------------------------------------------------------------------------

	In quantum physics, the study of physical systems in interaction consists of solving the Schr\"{o}dinger, Klein--Gordon, and Dirac equations. The solutions of these equations provide the energy spectra and wave functions that contain the full information associated with the investigated system. From a mathematical point of view, these equations are partial differential equations that are often reduced, under suitable physical assumptions, to second-order differential equations. The final form of the resulting equation depends on the interaction potential and on the mathematical model used to describe it. In recent years, many authors have investigated exact and approximate solutions of the Schr\"{o}dinger equation for different potential models and by using different mathematical methods.
	
	Exponential-type potentials have received particular attention because they are useful in molecular systems, as in the work of Malli on Hulth\'{e}n-type molecular integrals \cite{Malli}, and in different screened-interaction models studied by Myhrman for angular-momentum states \cite{Myhrman}. Related exponential-type structures also appear in atomic and quantum-mechanical models, such as the screened and multi-component systems considered by Arai \cite{Arai1}. In addition, several classical central potentials have played an important role in the literature: the Hulth\'{e}n potential \cite{Hulthen}, the Yukawa potential \cite{Yukawa}, the Kratzer potential \cite{Kratzer}, the Woods--Saxon potential \cite{Saxon}, and the Hellmann potential \cite{Hellman}. These models are widely used because they provide realistic descriptions of short-range interactions in different areas of physics.
	
	Since these potentials are central and spherically symmetric, the centrifugal term appears in the radial Schr\"{o}dinger equation \cite{Sadoun3} and in the Klein--Gordon equation \cite{Sadoun5}, \cite{Sadoun4}. This term makes it difficult to obtain exact analytical solutions for nonzero angular momentum. To overcome this difficulty, several approximations of the centrifugal term have been introduced so that it can be written in a form compatible with the potential under study. For example, Falaye and coauthors used the Nikiforov--Uvarov method for exact solutions of the Schr\"{o}dinger equation with the Hulth\'{e}n potential \cite{Falaye1}, while Ahmadov studied the same type of problem using supersymmetric quantum mechanics \cite{Ahmadov}. Other approaches include the Laplace transformation method \cite{Miraboutalebi}, the asymptotic iteration method \cite{Falaye2}, the series expansion method \cite{Ibekwe}, and the path integral method. In particular, path-integral treatments have been applied to the radial Schr\"{o}dinger equation for diatomic systems and related potentials by Sadoun et al. \cite{Sadoun1,Sadoun2}, and Guechi et al. \cite{Guechi2,Sadoun3}.
	
	Recently, many authors have studied approximate bound-state solutions of the Schr\"{o}dinger equation for linear combinations of known potentials. Onyenegecha et al. investigated the inversely quadratic Hellmann--Kratzer potential and analyzed its bound-state structure \cite{Onyenegecha1}. William and coauthors studied the Hulth\'{e}n--Hellmann combination \cite{William}, while Ita et al. considered the Manning--Rosen plus Hellmann potential \cite{Ita}. Other examples include the Hua plus modified Eckart potential \cite{Onyenegecha2}, the modified M\"{o}bius square plus Hulth\'{e}n potential \cite{Ukewuihe}, the modified M\"{o}bius square plus Kratzer potential \cite{Onyenegecha3}, the $q$-deformed Hulth\'{e}n plus generalized inverse quadratic Yukawa potential \cite{Edet}, and the Hulth\'{e}n-screened Kratzer potential \cite{Inyang}. These works show that combining potentials can produce richer spectral properties than a single potential model.
	
	Thermodynamic properties have also been studied for several specific potential models. In the non-relativistic Schr\"{o}dinger framework, Okorie et al. investigated the improved deformed exponential-type potential (IDEP) for diatomic molecules \cite{thermodynamic2020}. Ikot et al. studied the thermodynamics of diatomic molecules with a general molecular potential \cite{thermodynamic2018}. Louis et al. analyzed the Manning-Rosen plus Hellmann potential and its thermodynamic properties using the proper quantization rule \cite{thermodynamic2019}. Okorie et al. also examined the quadratic exponential-type potential in $D$-dimensions \cite{thermodynamic2018-2} and the modified Yukawa potential \cite{thermodynamic2018-3}. More recently, Inyang et al. studied the modified screened Kratzer plus inversely quadratic Yukawa potential and discussed thermodynamic properties for selected diatomic molecules \cite{thermodynamic2022}. In the relativistic setting, Onyeaju et al. obtained approximate bound-state solutions of the Dirac equation for the deformed Hylleraas plus deformed Woods-Saxon potential and derived the corresponding thermodynamic properties \cite{thermodynamic2017}. Demirci and Sever studied the Klein-Gordon equation with the Eckart plus a class of Yukawa potential and also derived its non-relativistic thermal properties \cite{thermodynamic2023}. 
	
	These studies motivated the present work, in which we investigate approximate analytical solutions using the Feynman path integral approach \cite{Feynman,Sadoun1,Sadoun2,Guechi1,Diaf1}, together with the thermodynamic properties of a non-relativistic quantum system under a linear combination of the generalized four-parameter potential and the Yukawa potential, given by
	
	\begin{equation}
		V\left( r\right) =\frac{a}{\left( e^{2\alpha r}-q\right) ^{2}}-\frac{b}{%
			e^{2\alpha r}-q}-\frac{ce^{-\alpha r}}{r},  \label{a.1}
	\end{equation}%
	where $a$, $b$, and $c$ are positive constants defined by $a=D_e  (e^{\alpha r_e}-1)^2$, $b=2D_e  (e^{\alpha r_e}-1)$, and $c=V_0$, where $D_e$ is the depth of the potential well and $r_e$ is the equilibrium distance of the two nuclei. The parameters $q$ and $\alpha$ are the deformation parameter and the screening parameter, respectively.
	
	This paper is organized as follows. In Sec. \ref{Sec:PIF}, we briefly present the path integral formalism for this model where we obtain the radial Green's function for the generalized four-parameter potential and the Yukawa potential. Then, we derive the exact energy spectrum and the normalized wave function for this potential in the diatomic-molecule case. In final subsection we exhibit the limit cases of the energy spectrum. In Sec. \ref{Sec:TP}, the explicit expressions of the thermodynamic properties are obtained, and the results are discussed in Sec. \ref{Sec:RD}. Finally, we present our conclusions.
	%---------------------------------------------------------------------------------------
	\section{Path integral formalism and energy spectrum}\label{Sec:PIF}
	%---------------------------------------------------------------------------------------
	
	Within the framework of path integral formalism, the Hamiltonian of a non-relativistic spinless system subjected to central potential with spherical symmetry (\ref{a.1}); is written as follow
	\begin{equation}
		H_{l}=\frac{P_{r}^{2}}{2m}+V\left( r\right) +\frac{\hbar ^{2}l\left(
			l+1\right) }{2mr^{2}}.  \label{a.4}
	\end{equation}
	
	\begin{figure}[ht]
		\centering
		\includegraphics[width=0.6\textwidth]{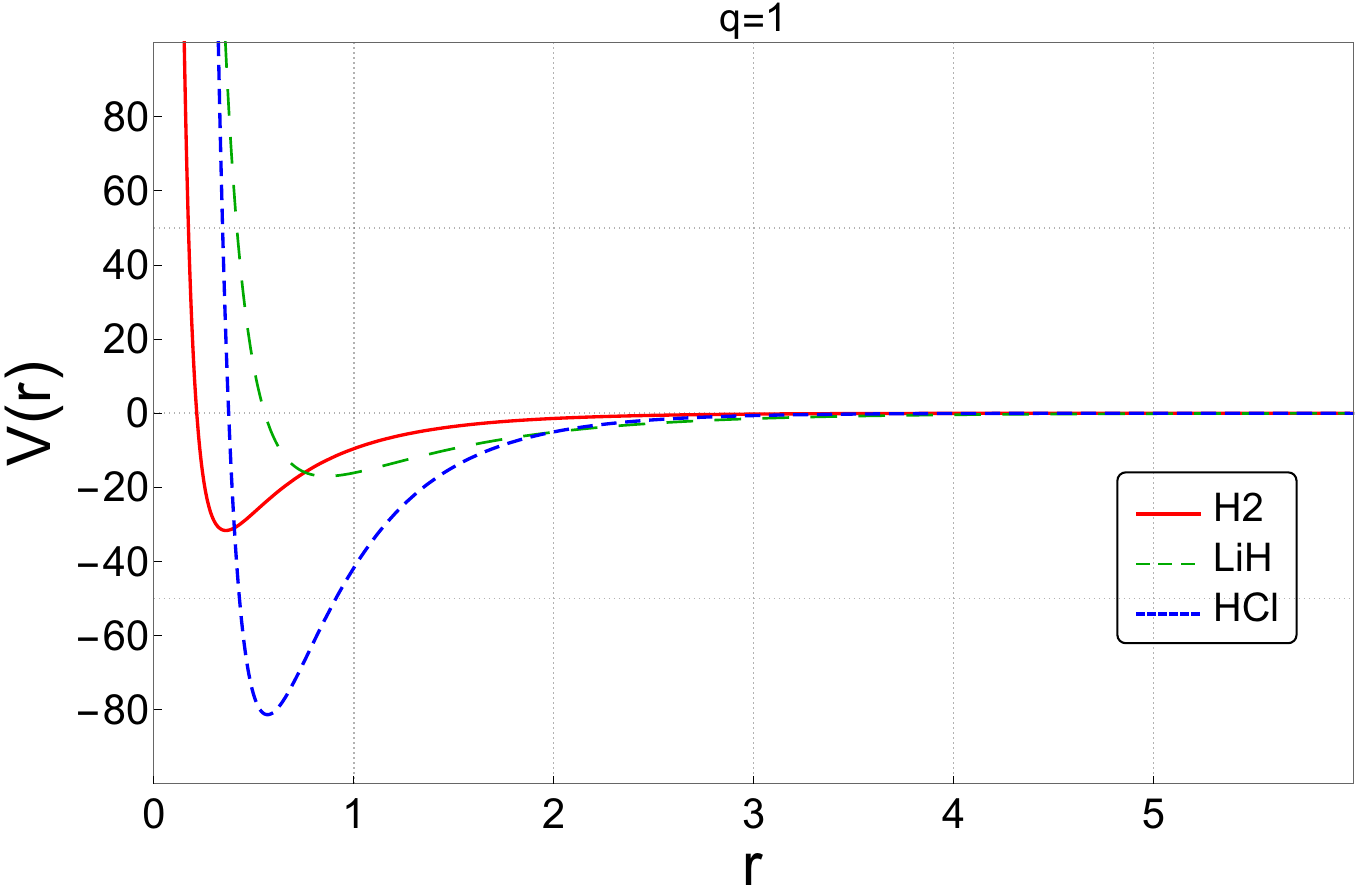}
		\caption{The variation of linear combination of the generalized four-parameter potential and the Yukawa potential as a function of $r$, for selected diatomic molecules.}  \label{fig:V1}
	\end{figure}
	The Fig. \ref{fig:V1}, illustrate the behavior of the linear combination of the generalized four-parameter potential and the Yukawa potential \eqref{a.1} as a function of $r$ for some diatomic molecules, with different band type. It is clear that, this combination of potentials exhibit a minimum, that can describe stable region in which the molecule can have vibrational stat. Which represent an advantage to study thermodynamical properties for such molecules in their vibrational states.
	
		\subsection{Propagator construction and Green's function}
	The first step of the investigation of such potential consists in writing  the Green's function solution of Schr\"odinger equation which reads:
	\begin{equation}
		G\left( \vec{r}^{\prime \prime },\vec{r}^{\prime };E\right) =\frac{1}{%
			r"r^{\prime }}\sum\limits_{l=0}^{\infty }\frac{2l+1}{4\pi }G_{l}\left(
		r^{\prime \prime },r^{\prime },E\right) P_{l}\left( \cos \theta \right)
		\label{a.2}
	\end{equation}%
	where the Legendre polynomial describes the angular part $P_{l}\left( \cos \theta \right)$ with $\cos \theta =\left( \vec{r}^{\prime \prime },\vec{r}^{\prime }\right)$. The radial part is expressed in term of the radial Green's function $G_{l}\left( r^{\prime \prime },r^{\prime },E\right)$ given by \cite{Grosche1}:
	\begin{equation}
		G_{l}\left( r^{\prime \prime },r^{\prime },E\right) =\frac{i}{\hbar }%
		\int\limits_{0}^{\infty }dT\left\langle r^{\prime \prime }\right\vert \exp %
		\left[ -\frac{i}{\hbar }T\left( H_{l}-E\right) \right] \left\vert r^{\prime
		}\right\rangle  \label{a.3}
	\end{equation}%
	
	The obtaining energy spectrum and normalized wave function involves calculating Green's function
	(\ref{a.3}). To calculate Green's function, we need to use approximations to circumvent the difficulty of obtaining an exact analytical solution. Such a solution is impossible to obtain because of the difference between the behavior of the functions that appear in the expression of the potential (\ref{a.1}) and the centrifugal term.
	
	We use the following approximations which contains the deformed parameter $q$ and screening parameter $\alpha$ \cite{greene1,Sadoun3}:
	\begin{equation}
		\dfrac{1}{r}\approx \dfrac{2\alpha e^{-\alpha r}}{1-qe^{-2\alpha r}},\quad\, \text{and}\quad  
		\dfrac{1}{r^{2}}\approx \dfrac{4\alpha ^{2}e^{-2\alpha r}}{\left(
			1-qe^{-2\alpha r}\right) ^{2}}  \label{a.5}
	\end{equation}
	The validity of this approximation goes only for $q\geq 1$ (for more detail see Refs. \cite{Sadoun3}). Substituting inside the Hamiltonian (\ref{a.4}) gives the new form:
	\begin{equation}
		H_{l}=\frac{P_{r}^{2}}{2m}-\frac{\left( b+2\alpha c\right)
			e^{-2\alpha r}}{1-qe^{-2\alpha r}}+\frac{ae^{-4\alpha r}}{%
			\left( 1-qe^{-2\alpha r}\right) ^{2}} +\frac{4\alpha ^{2}\hbar ^{2}l\left( l+1\right) e^{-2\alpha r}}{2m\left(
			1-qe^{-2\alpha r}\right) ^{2}}  \label{a.6}
	\end{equation}%
	In present form the above Hamiltonian present a strong singularity at $r_{0}=\frac{1}{2\alpha }\ln q$, which gives rise to two different regions $] 0,r_{0}[ $ and $] r_{0},+\infty [ $. The first region is not of great physical interest since it describes a particle confined within a sphere of radius $r_0$; an analytical solution in this case is impossible. Our interest lies in the second region, where we write the expression (\ref{a.3}) as a path integral
	\begin{equation}
		G_{l}\left( r^{\prime \prime },r^{\prime },E\right) =\frac{i}{\hbar }%
		\int\limits_{0}^{\infty }dSK_{l}\left( r^{\prime \prime },r^{\prime
		};S\right) ,  \label{a.7}
	\end{equation}%
	
	and we introduce a regulating function $f\left( r\right) $ \cite{Kleinert} defined by
	\begin{equation}
		f\left( r\right) =f_{R}\left( r\right) f_{L}\left( r\right) =f^{1-\lambda
		}\left( r\right) f^{\lambda }\left( r\right) .  \label{a.12}
	\end{equation}
	in order to obtain a discrete form of the path integral. Using the function $f\left( r\right) $ we write the propagator $K_{l}\left( r^{\prime \prime },r^{\prime};S\right)$ in the equation (\ref{a.7}) as
	
	\begin{align}
		K_{l}\left( r^{\prime \prime },r^{\prime };S\right) &=
		f_{R}\left( r^{\prime
			\prime }\right) f_{L}\left( r^{\prime }\right) \left\langle r^{\prime \prime
		}\right\vert \exp \left[ -\frac{i}{\hbar }S\left( \frac{P_{r}^{2}}{2m}%
		+  V\left( r\right) +\frac{\hbar ^{2}l\left( l+1\right) }{%
			2mr^{2}}-E\right) \right] \left\vert r^{\prime }\right\rangle  \notag \\
		&=f_{R}\left( r^{\prime \prime }\right) f_{L}\left( r^{\prime }\right) \int
		Dr\left( s\right) \int \frac{DP_{r}\left( s\right) }{2\pi \hbar }\exp
		\left\{ \frac{i}{\hbar }\int\limits_{0}^{S}\left[ P_{r}\dot{r}- f_{L}\left( r\right) \left( \frac{P_{r}^{2}}{2m}+V\left(
		r\right) +\frac{\hbar ^{2}l\left( l+1\right) }{2mr^{2}}-E\right) f_{R}\left(
		r\right) \right] ds\right\} .  \label{a.8}
	\end{align}
	
	In a discrete form%
	\begin{equation}
		K_{l}\left( r^{\prime \prime },r^{\prime };S\right) =f_{R}\left( r^{\prime
			\prime }\right) f_{L}\left( r^{\prime }\right) \lim\limits_{N\rightarrow
			\infty }\prod_{j=1}^{N}\left[ \int dr_{j}\right] \prod_{j=1}^{N+1}\left[
		\int \frac{d\left( P_{r}\right) _{j}}{\left( 2\pi \hbar\right) }\right] 
		\exp \left[ \frac{i}{\hbar}\sum_{j=1}^{N+1}\mathcal{A}_{1}^{j}\right] ,
		\label{a.9}
	\end{equation}%
	with the action $\mathcal{A}_{1}^{j}$ given by%
	\begin{equation}
		\mathcal{A}_{1}^{j}=-\left( P_{r}\right) _{j}\Delta r_{j}-\varepsilon
		_{s}f_{L}\left( r_{j}\right) \left[ \frac{\left( P_{r}\right) _{j}^{2}}{2m}%
		+V_{eff}\left( r_{j}\right) -E\right] f_{R}\left( r_{j-1}\right)
		\label{a.10}
	\end{equation}%
	and 
	\begin{equation}
		\varepsilon _{s}=\frac{S}{N+1}=ds=\frac{dt}{f_{L}\left(
			r_{j}\right) f_{R}\left( r_{j-1}\right) },\text{ \ }dt=\varepsilon _{t}=%
		\frac{T}{N+1}.  \label{a.11}
	\end{equation}
	The choice of $\lambda =\frac{1}{2}$ corresponds to the prescription of the mid-point. 
	This value allows us to make a development of the action around the mid-point, thus simplify the integration on the variables $\left( P_{r}\right) _{j}$, the propagator become
	
	\begin{equation}
		K_{l}\left( r^{\prime \prime },r^{\prime };S\right) =\left[ f\left(
		r^{\prime \prime }\right) f\left( r^{\prime }\right) \right] ^{\frac{1}{4%
		}}\lim\limits_{N\rightarrow \infty }\prod_{j=1}^{N+1}\left[ \frac{m}{2i\pi
			\hbar \varepsilon _{s}}\right] ^{\frac{1}{2}}   \prod_{j=1}^{N}\left[ \int \frac{dr_{j}}{\sqrt{f\left( r_{j}\right) 
		}}\right] \exp \left\{ \frac{i}{\hbar }\sum_{j=1}^{N+1}\mathcal{A}%
		_{2}^{j}\right\}  \label{a.13}
	\end{equation}%
	with the action
	\begin{equation}
		\mathcal{A}_{2}^{j}=\frac{m\left( \Delta r_{j}\right) ^{2}}{2\varepsilon _{s}%
			\sqrt{f\left( r_{j}\right) f\left( r_{j-1}\right) }}-\varepsilon _{s}\left(
		V_{eff}(r_{j})-E\right) \sqrt{f\left( r_{j}\right) f\left( r_{j-1}\right) }
		\label{a.14}
	\end{equation}%
	and effective potential%
	\begin{equation}
		V_{eff}\left( r\right) =\frac{4\alpha ^{2}\hbar ^{2}l\left( l+1\right)
			e^{-2\alpha r}}{2m\left( 1-qe^{-2\alpha r}\right) ^{2}}-\frac{\left(
			b+2\alpha c\right) e^{-2\alpha r}}{1-qe^{-2\alpha r}}+\frac{ae^{-4\alpha r}}{\left( 1-qe^{-2\alpha r}\right) ^{2}}.  \label{a.15}
	\end{equation}

	%---------------------------------------------------------------------------------------%---------------------------------------------------------------------------------------

	%---------------------------------------------------------------------------------------
%	\section{Evaluation of the radial Green's function}\label{Sec:GF}
	%---------------------------------------------------------------------------------------
	
	As first step, one has to remove the singularity at $r_{0}=\frac{1}{2\alpha }\ln q$ present in the effective potential \eqref{a.15}. To realize that, we apply a spatial transformation $r \rightarrow \xi$
	\begin{equation}
		r=\frac{1}{2\alpha }\ln \left( \exp \left( 4\alpha \xi \right) +q\right)=h(\xi),
		\label{a.16}
	\end{equation}%
	where for $r\in [r_0,+\infty[$ we have $\xi \in ]-\infty,+\infty]$. Then, we use the deformed hyperbolic functions introduced by Arai \cite{Arai1,Arai2}
	\begin{equation}
		\left\{ 
		\begin{array}{c}
			\sinh _{q}x=\frac{e^{x}-qe^{-x}}{2},\cosh _{q}x=\frac{e^{x}+qe^{-x}}{2}%
			,\tanh _{q}x=\frac{\sinh _{q}x}{\cosh _{q}x}, \\ 
			e^{2x}=2\cosh _{q}^{2}x+2\cosh _{q}x\sinh _{q}x-q,\,\,\qquad\qquad \\ 
			e^{-2x}=\frac{1}{q^2}\left( 2\cosh _{q}^{2}x-2\cosh _{q}x\sinh _{q}x-q\right) ,\,\,\,\,\,\,\quad\quad%
		\end{array}%
		\right.  \label{a.19}
	\end{equation}
	In this case, the regulating function reads%
	\begin{equation}
		f\left[ r\left( \xi \right) \right] =\frac{\exp \left( 4\alpha \xi \right) }{%
			\cosh _{q }^{2}\left( 2\alpha \xi \right) }=\left[ h^{\prime }\left(
		\xi \right) \right] ^{2},  \label{a.17}
	\end{equation}
	
	and the propagator of nonzero angular momentum ($l \neq 0$) given by equation (\ref{a.13}) in its new form is written as
	\begin{eqnarray}
		K_{l}\left( r^{\prime \prime },r^{\prime };S^{\prime }\right) &=&\left[
		f\left( r^{\prime \prime }\right) f\left( r^{\prime }\right) \right] ^{\frac{%
				1}{4}}\lim\limits_{N\rightarrow \infty }\prod_{n=1}^{N+1}\sqrt{\frac{m}{%
				2i\pi \varepsilon _{s}}}\prod_{j=1}^{N}\left[ \int d\xi _{j}\right]
		\exp \left\{ \frac{i}{\hbar}\sum_{j=1}^{N+1}\left[ m\frac{\left( \Delta \xi
			_{j}\right) ^{2}}{2\varepsilon _{s}}+\frac{m}{8\varepsilon _{s}}%
		\left( \left( \frac{h^{\prime \prime }}{h^{\prime }}\right) ^{2}-\frac{2}{3}%
		\frac{h^{\prime \prime \prime }}{h^{\prime }}\right) \left( \Delta \xi
		_{j}\right) ^{4}\right. \right.  \notag \\
		&&+\varepsilon _{s}\left( 2E-\frac{2}{q^{2}}\left[ \frac{4q\alpha ^{2}\hbar
			^{2}l\left( l+1\right) }{2m}+a\right] \right) +\varepsilon _{s}\left( 2E+\frac{2}{q^{2}}\left[ \frac{4q\alpha ^{2}\hbar
			^{2}l\left( l+1\right) }{2m}+a\right] \right) \tanh
		_{q}\left( 2\alpha \xi _{j}\right)  \notag \\
		&&+\left. \left. \varepsilon _{s}\left( b+2\alpha c+\frac{a}{q}-qE\right) \frac{1}{\cosh _{q}^{2}\left( \alpha \xi _{j}\right) 
		}\right] \right\} .  \label{a.18}
	\end{eqnarray}
	Notice that perturbation theory allows us to estimate the term in $\left( \Delta \xi _{j}\right) ^{4}$ appearing in the kernel (\ref{a.18}) as follows
	\begin{eqnarray}
		\left\langle \left( \triangle \xi _{n}\right) ^{4}\right\rangle
		&=&\int_{-\infty }^{+\infty }d\left( \triangle \xi _{n}\right) \left(
		\triangle \xi _{n}\right) ^{4}\left[ \frac{m}{2i\pi \hbar \varepsilon _{s}}%
		\right] ^{\frac{1}{2}}\exp \left[ \frac{im}{2\hbar \varepsilon _{s}}\left(
		\triangle \xi _{n}\right) ^{2}\right]  \notag \\
		&=&-3\varepsilon _{s}^2\left( \frac{\hbar }{m}\right) ^{2}.  \label{a.20}
	\end{eqnarray}
	This term contributes significantly to the path integral and for more detail see Ref. \cite{Sadoun2}.

	Finally, we put the radial Green's function (\ref{a.7}) in its new form using the change of variables $\xi =\frac{y}{2\alpha }\rightarrow u=y-\frac{1}{2}\ln q,$ $4\alpha ^{2}ds=d\tau $ and\ \ $4\alpha ^{2}S=\Lambda $. One finds
	\begin{eqnarray}
		G_{l}\left( r^{\prime \prime },r^{\prime };E\right) &=&\left[ f\left(
		r^{\prime \prime }\right) f\left( r^{\prime }\right) \right] ^{\frac{1}{4}%
		}\int_{0}^{\infty }d\Lambda \text{ }\exp \left( \frac{i}{\hbar }%
		E_{RM}\Lambda \right) K_{RM}\left( u^{\prime \prime },u^{\prime };\Lambda
		\right)  \notag \\
		&=&\left[ f\left( r^{\prime \prime }\right) f\left( r^{\prime }\right) %
		\right] ^{\frac{1}{4}}G_{RM}\left( u^{\prime \prime },u^{\prime };E\right) .
		\label{a.21}
	\end{eqnarray}%
	In above equation the propagator $K_{RM}\left( u^{\prime \prime },u^{\prime };\Lambda \right)$ is the one relating to the Rosen-Morse potential \cite{Diaf2,Guechi2} (general modified Poschl--Teller potential)
	\begin{eqnarray}
		K_{RM}\left( u^{\prime \prime },u^{\prime };\Lambda \right) &=&\int 
		\mathfrak{D}u\left( \tau \right) \exp \left\{ \frac{i}{\hbar }%
		\int\limits_{0}^{\Lambda }d\tau \left[ \frac{m}{2}\dot{u}^{2}\right. \right.
		+ \frac{1}{4\alpha^2}\left( E+\frac{1}{q^{2}}\left[ \frac{4q\alpha ^{2}\hbar ^{2}l\left(
			l+1\right) }{2m}+a\right] +\alpha ^{2}\frac{\hbar ^{2}}{2m}%
		\right) \tanh u  \notag \\
		&&+\left. \left. \frac{1}{4\alpha^2}\left( \frac{b+2\alpha c}{q}+\frac{a%
		}{q^{2}}-E-\alpha ^{2}\frac{\hbar ^{2}}{2m}\right) \frac{1}{\cosh ^{2}u}%
		\right] \right\}  \notag \\
		&=&\int \mathcal{D}u\left( \tau \right) \exp \left\{
		i\int\limits_{0}^{\Lambda }\left[ \frac{\dot{u}^{2}}{2}-V_{RM}^{l}\left(
		u\right) \right] d\tau \right\} ,  \label{a.23}
	\end{eqnarray}
	with the Rosen-Morse potential defined in terms of deformed hyperbolic functions
	\begin{equation}
		V_{RM}^{l}\left( u\right) =A_{l}\tanh u-\frac{B}{\cosh ^{2}u},\text{ \ \ \ \ 
		}u\in \mathcal{R},  \label{a.24}
	\end{equation}%
	where we put%
	\begin{equation}
		\left\{ 
		\begin{array}{l}
			A_{l}=-\frac{1}{2\alpha ^{2}}\left( E+\frac{1}{q^{2}}\left[ \frac{4q\alpha
				^{2}\hbar ^{2}l\left( l+1\right) }{2m}+a\right] +\alpha ^{2}%
			\frac{\hbar ^{2}}{2m}\right) , \\ 
			B=\frac{1}{4\alpha ^{2}}\left( \frac{b+2\alpha c}{q}+\frac{a}{q^{2}}-E-\alpha ^{2}\frac{\hbar ^{2}}{2m}\right) .%
		\end{array}%
		\right.  \label{a.25}
	\end{equation}
	The corresponding energy relating to the Rosen-Morse potential is given by 
	\begin{equation}
		E_{RM}=\frac{1}{2\alpha ^{2}}\left( E-\frac{1}{q^{2}}\left[ \frac{4q\alpha
			^{2}\hbar ^{2}l\left( l+1\right) }{2m}+a\right] -\alpha ^{2}%
		\frac{\hbar ^{2}}{2m}\right) . \label{a.22}
	\end{equation}%
	
	We can now write down the explicit expression for the Green's function using the well known exact solution \cite{Grosche1},
	\begin{equation}
		G_{RM}\left( u^{\prime \prime },u^{\prime };E\right) =\frac{m}{i\hbar }%
		\Gamma \left( M_{1}-L_{E}\right) \Gamma \left( L_{E}-M_{1}+1\right)  d_{M_{1},M_{2}}^{L_{E}}\left( \theta ^{\prime \prime }-\pi \right)
		d_{M_{1},M_{2}}^{L_{E}\ast }\left( \theta ^{\prime }\right) \text{ },
		\label{a.26}
	\end{equation}%
	here, $d_{M_{1},M_{2}}^{L_{E}}\left( \theta \right) $ is the Wigner function, with $\tanh u=-\cos \theta ,\theta \in \left( 0,\pi \right) .$ The indices $L_{E},M_{1}$ and $M_{2}$ are defined by \cite{Grosche1}:
	\begin{align}
		L_{E} &=L_{B}=-\frac{1}{2}+\frac{1}{2}\sqrt{\frac{8mB}{\hbar^2}+1} 
		\notag \\
		&=-\frac{1}{2}+\frac{1}{2}\sqrt{\frac{2m}{\hbar ^{2}} \left( \frac{b+2\alpha c}{\alpha ^{2}q}+\frac{a}{\alpha^2q^{2}}\right) -\frac{2mE}{\alpha ^{2}\hbar ^{2}}},  \label{a.27}
	\end{align}
	and
	\begin{eqnarray}
		M_{1/2} =\sqrt{\frac{m}{2\hbar ^{2}}}\left( \sqrt{-A_{l}-E_{RM}}\pm\sqrt{%
			A_{l}-E_{RM}}\right)  =\sqrt{\frac{2m}{q^{2}\hbar ^{2}}\left( \frac{q\hbar ^{2}l\left(
				l+1\right) }{2m}+\frac{a}{4\alpha^2}\right) +\frac{1}{4}}\pm\frac{1}{2}\sqrt{-\frac{2mE}{%
				\alpha ^{2}\hbar ^{2}}},  \label{a.28}
	\end{eqnarray}%
	%\begin{eqnarray}
	%	M_{2} &=&\sqrt{\frac{m}{2\hbar ^{2}}}\left( \sqrt{-A_{l}-E_{RM}}-\sqrt{%
		%		A_{l}-E_{RM}}\right)  \notag \\
	%	&=&\sqrt{\frac{2m}{q^{2}\hbar ^{2}}\left( \frac{q\hbar ^{2}l\left(
			%			l+1\right) }{2m}+\frac{a}{4\alpha^2}\right) +\frac{1}{4}}-\frac{1}{2}\sqrt{-\frac{2mE}{%
			%			\alpha ^{2}\hbar ^{2}}}.  \label{a.29}
	%\end{eqnarray}

	%-----------------------------------------------------------------------------------------------%-----------------------------------------------------------------------------------------------

	%-----------------------------------------------------------------------------------------------
	\subsection{Energy spectrum and wave functions}\label{Sec:ESWF}
	%-----------------------------------------------------------------------------------------------
	
	The analysis of the Green's function poles occurring in the Euler function $\Gamma\left( M_{1}-L_{E}\right) $ when $M_{1}-L_{E}=-n$, for $n=0,1,2,..$ gives the discrete energy spectrum
	
	\begin{equation}
		E_{n,l}=-\frac{\hbar^2}{2m}\frac{1}{q^{2}}\left[ \frac{\frac{\alpha
				^{2}q^{2}m}{2\hbar ^{2}}\left( \frac{b+2\alpha c}{\alpha ^{2}q}+%
			\frac{a}{\alpha^2 q^{2}}\right) -\left( \alpha qn+N_{l}\right) ^{2}}{\alpha
			qn+N_{l}}\right] ^{2},  \label{a.30}
	\end{equation}%
	where%
	\begin{equation}
		N_{l}=\frac{\alpha q}{2}\left( 1+\sqrt{1+\frac{4l\left( l+1\right) }{q}+\frac{%
				2m}{\hbar ^{2}}\left( \frac{a}{\alpha^2 q^{2}}\right)}\right) .  \label{a.31}
	\end{equation}
	
	Now we have to express the discrete part of Green's function $G_{l}\left( r^{\prime\prime },r^{\prime };E\right) $ as a spectral expansion. For this, we first use the approximation of Euler function $\Gamma\left(M_{1}-L_{E}\right)$ in vicinity of the poles $M_{1}-L_{E}=-n$,  \cite{Mustapic}:
	\begin{eqnarray}
		\Gamma \left( M_{1}-L_{E}\right) &\approx &\frac{\left( -1\right) ^{n}}{n!}%
		\frac{1}{M_{1}-L_{E}+n}  \notag \\
		&=&\frac{\left( -1\right) ^{n}}{n!}\frac{\alpha \hbar ^{2}}{m}\frac{%
			Q_{l}\left( 2(N_{l}+\alpha qn)-qQ_{l}\right) }{\left( \alpha qn+N_{l}\right)
			\left( E-E_{n,l}\right) },  \label{a.32}
	\end{eqnarray}%
	with%
	\begin{equation}
		Q_{l}=\frac{1}{q}\frac{\left( \alpha qn+N_{l}\right) ^{2}-\frac{\alpha
				^{2}q^{2}m}{2\hbar ^{2}}\left( \frac{b+2\alpha c}{\alpha ^{2}q}+%
			\frac{a}{\alpha^2 q^{2}}\right) }{\alpha qn+N_{l}}.  \label{a.33}
	\end{equation}%
	then we use the Wigner's function symmetry property%
	\begin{equation}
		d_{M_{1},M_{2}}^{L_{E}}\left( \theta \right) =\left( -1\right)
		^{L_{E}-M_{1}}d_{M_{1},-M_{2}}^{L_{E}}\left( \theta -\pi \right).
		\label{a.34}
	\end{equation}%
	We find in the end:
	\begin{eqnarray}
		G_{l}\left( r^{\prime \prime },r^{\prime };E\right) &=&-i\hbar \frac{\left[
			f\left( r^{\prime \prime }\right) f\left( r^{\prime }\right) \right] ^{\frac{%
					1}{4}}}{r^{\prime \prime }r^{\prime }}\sum_{n=1}^{n_{\max }}\eta \frac{%
			Q_{l}\left( n+N_{l}-Q_{l}\right) }{n!\left( n+N_{l}\right) \left(
			E-E_{n,l}\right) }  d_{-\frac{1}{2}+\frac{N_{l}}{\alpha q}-\frac{Q_{l}}{\alpha },\frac{1%
			}{2}-\frac{N_{l}}{\alpha q}-\frac{Q_{l}}{\alpha }}^{-\frac{1}{2}+n+\frac{%
				N_{l}}{\alpha q}-\frac{Q_{l}}{\alpha }}\left( \theta ^{\prime \prime
		}\right) d_{-\frac{1}{2}+\frac{N_{l}}{\alpha q}-\frac{Q_{l}}{\alpha q},\frac{%
				1}{2}-\frac{N_{l}}{\alpha q}-\frac{Q_{l}}{\alpha }}^{-\frac{1}{2}+n+\frac{%
				N_{l}}{\alpha q}-\frac{Q_{l}}{\alpha }}\left( \theta ^{\prime }\right) 
		\notag \\
		&=&i\hbar \sum_{n=1}^{n_{\max }}\frac{\chi _{n,l}^{q\geq 1\ast }\left(
			r^{\prime }\right) \chi _{n,l}^{q\geq 1}\left( r^{\prime \prime }\right) }{%
			E-E_{n,l}^{q\geq 1}},  \label{a.35}
	\end{eqnarray}%
	where $\chi _{n,l}^{\lambda \geq 1}\left( r\right)$, describe the normalized wave functions%
	\begin{equation}
		\chi _{n,l}\left( r\right) =\left[ -\frac{2Q_{l}\left(2( \alpha
			qn+N_{l})-qQ_{l}\right) }{n!\left( \alpha qn+N_{l}\right) }\right] ^{\frac{1}{%
				2}}\frac{f^{\frac{1}{4}}\left( r\right) }{r}d_{-\frac{1}{2}+\frac{N_{l}}{%
				\alpha q}-\frac{Q_{l}}{\alpha },\frac{1}{2}-\frac{N_{l}}{\alpha q}-\frac{%
				Q_{l}}{\alpha }}^{-\frac{1}{2}+n+\frac{N_{l}}{\alpha q}-\frac{Q_{l}}{\alpha }%
		}\left( \theta \right) .  \label{a.36}
	\end{equation}
	
	We introduce the hypergeometric functions using the following relation \cite{Gradshtein}
	\begin{eqnarray}
		d_{M_{1},M_{2}}^{L_{E}}\left( \theta \right) &=&\left[ \frac{\Gamma \left(
			L_{E}+M_{1}+1\right) \Gamma \left( L_{E}-M_{2}+1\right) }{\Gamma \left(
			L_{E}-M_{1}+1\right) \Gamma \left( L_{E}+M_{2}+1\right) }\right] ^{\frac{1}{2%
		}}  \frac{1}{\Gamma \left( M_{1}-M_{2}+1\right) }\left( \frac{1-\cos
			\theta }{2}\right) ^{\frac{M_{1}-M_{2}}{2}}\left( \frac{1+\cos \theta }{2}%
		\right) ^{\frac{M_{1}+M_{2}}{2}}  \notag \\
		&&\times _{2}F_{1}\left( -L_{E}+M_{1},L_{E}+M_{1}+1,M_{1}-M_{2}+1;\frac{%
			1-\cos \theta }{2}\right) ,  \notag \\
		&&  \label{a.37}
	\end{eqnarray}%
	finally the wave functions express as%
	\begin{eqnarray}
		\chi _{n,l}^{q\geq 1}\left( r\right) &=&\left[ \frac{-Q_{l}\left( 2(\alpha
			qn+N_{l})-qQ_{l}\right) }{\left( \alpha qn+N_{l}\right) }\frac{\Gamma \left(
			n+\frac{2N_{l}}{\alpha q}\right) \Gamma \left( n+\frac{2N_{l}}{\alpha q}-%
			\frac{2Q_{l}}{\alpha }\right) }{n!\Gamma \left( n-\frac{2Q_{l}}{\alpha }%
			+1\right) }\right] ^{\frac{1}{2}}   \frac{1}{\Gamma \left( \frac{2N_{l}}{\alpha q}\right) }\left(
		1-qe^{-2\alpha r}\right) ^{\frac{N_{l}}{\alpha q}}\left( qe^{-2\alpha
			r}\right) ^{-\frac{Q_{l}}{\alpha }}  \notag \\
		&&\times _{2}F_{1}\left( -n,n+\frac{2N_{l}}{\alpha q}-\frac{2Q_{l}}{\alpha },%
		\frac{2N_{l}}{\alpha q};1-qe^{-2\alpha r}\right) .  \label{a.38}
	\end{eqnarray}
	
	In order to ensure that the wave function $\chi _{n,l}^{q\geq 1}\left( r\right)$ remains finite as $r\rightarrow \infty $, we must impose the following condition $Q_{l}<0$. This yields:
	
	\begin{eqnarray}
		n_{\max } &=&\left\{ -\left( \frac{1}{2}+\frac{1}{2}\sqrt{\frac{4l\left(
				l+1\right) }{q}+\frac{2m}{\hbar ^{2}}\left( \frac{a}{\alpha^2 q^{2}}\right) +1}%
		\right) +\sqrt{\frac{\alpha ^{2}q^{2}m}{2\hbar ^{2}}%
			\left( \frac{b+2\alpha c}{\alpha ^{2}q}+\frac{a}{\alpha^2 q^{2}}\right) }%
		\right\}  \label{a.39}
	\end{eqnarray}

	\begin{figure}[ht]
		\centering
		\includegraphics[width=0.5\textwidth]{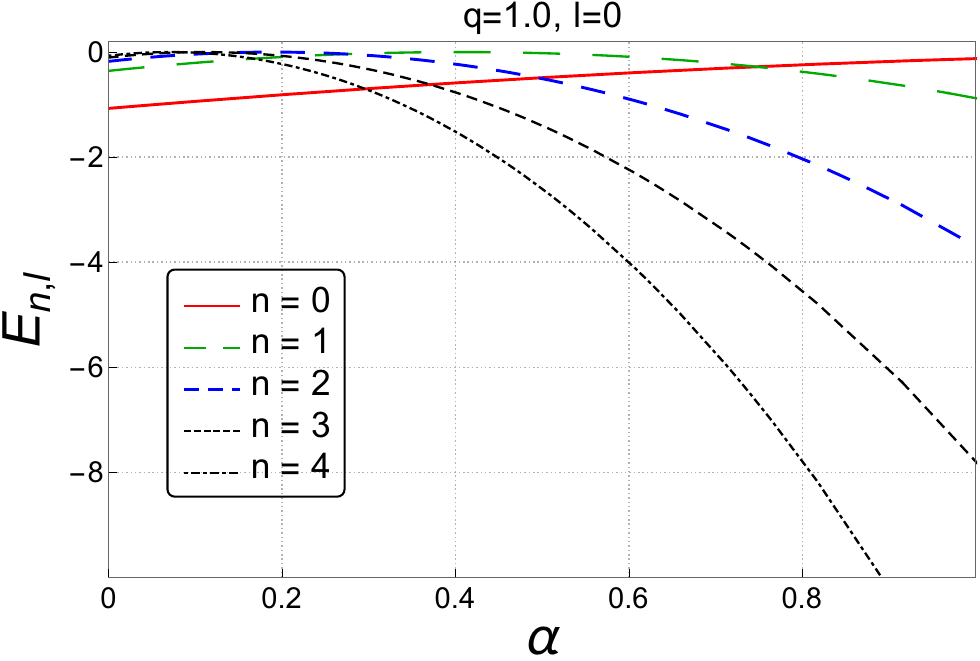}\hfill
		\includegraphics[width=0.5\textwidth]{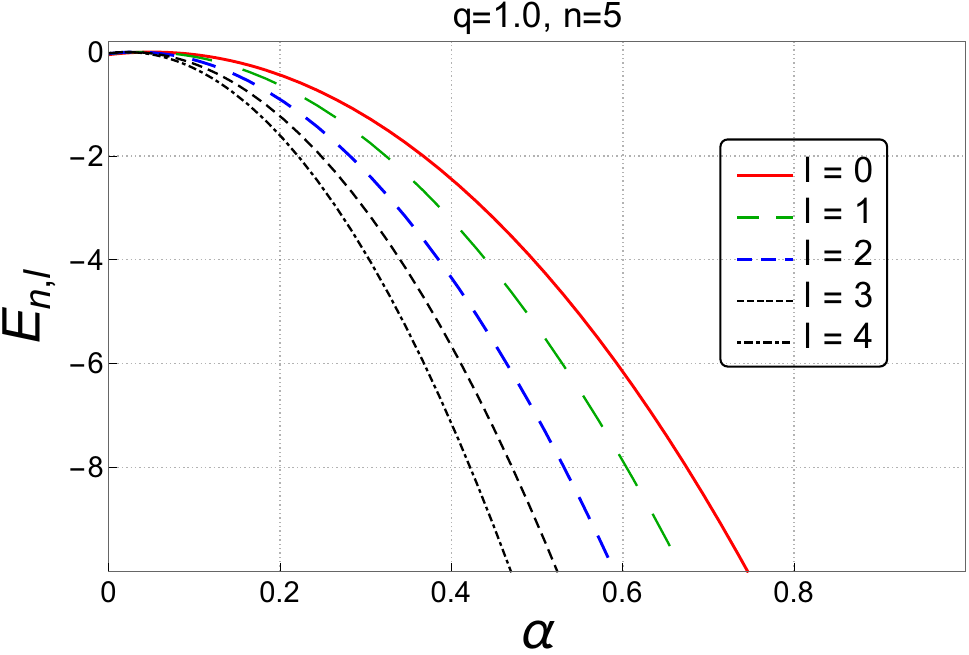}
		\caption{Variation of the energy against the screening parameter $\alpha$ for different quantum states: for $n$ at the left panel and for $l$ at the right panel.}  \label{fig01}
	\end{figure}
	\begin{figure}[ht]
		\centering
		\includegraphics[width=0.5\textwidth]{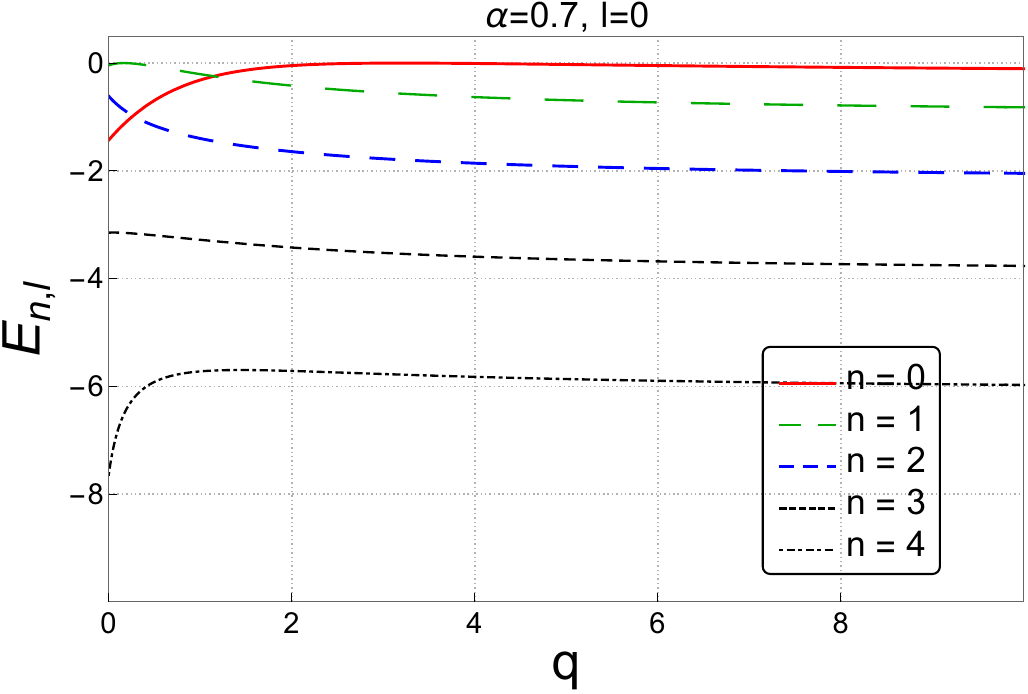}\hfill
		\includegraphics[width=0.5\textwidth]{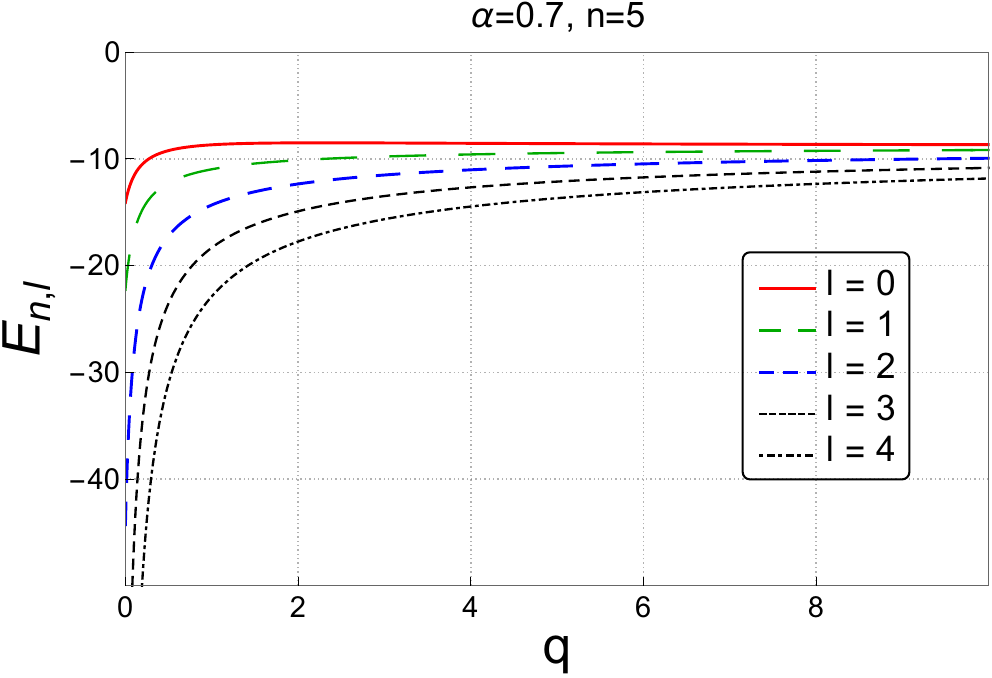}
		\caption{Variation of the energy according to the values of the deformation parameter $q$ for different quantum states: for $n$ at the left panel and for $l$ at the right panel.}  \label{fig02}
	\end{figure} 
	
	\begin{figure}[ht]
		\centering
		\includegraphics[width=0.5\textwidth]{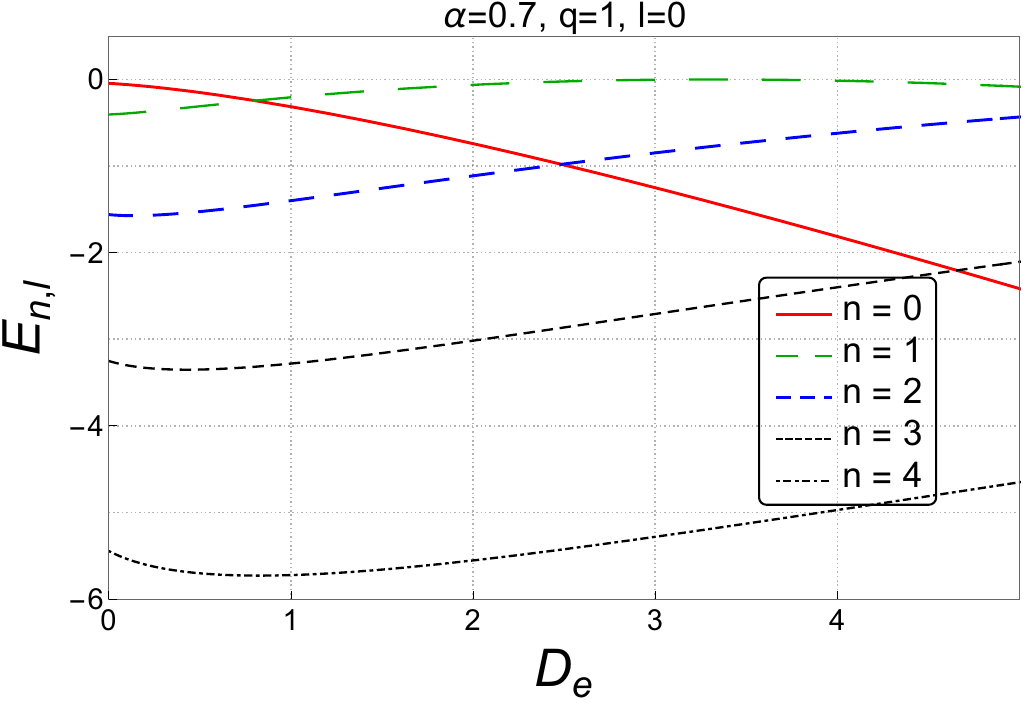}\hfill
		\includegraphics[width=0.5\textwidth]{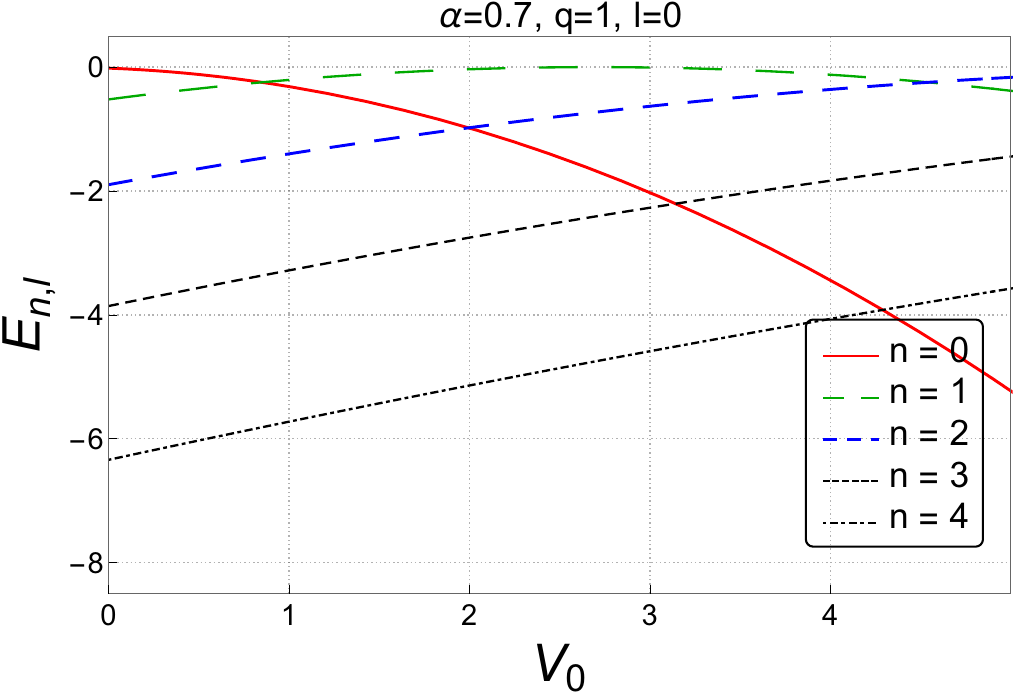}
		\caption{Variation of the energy as a function of the dissociation energy $D_e$ in the left panel and as a function of the depth of the potential well $V_0$ in the right panel, and for different quantum state $n$.}  \label{fig03}
	\end{figure}
	
	\subsection{Special cases for energy spectrum}
%---------------------------------------------------------------------------------------

So as to verify our results, we consider the following special cases.

%---------------------------------------------------------------------------------------
\subsubsection{Standard Hulth\'{e}n potential}
%---------------------------------------------------------------------------------------

By setting $a=0,c=0,q=1,2\alpha =\eta ,b=\eta Ze^{2}$, $Ze$ being the charge
of the nucleus, the energy spectrum expression for the standard Hulth\'{e}n
potential takes the form

\begin{equation}
	E_{n,l}=-\frac{\hbar ^{2}}{2m}\left[ \frac{\frac{m\eta Ze^{2}}{2\hbar ^{2}}%
		}{\frac{\eta n}{2}+N_{l}}-\left( \frac{\eta n}{2}+N_{l}\right)\right]
	^{2},
\end{equation}%
where%
\begin{equation}
	N_{l}=\frac{\eta }{2}\left(l+1\right) .
\end{equation}%

These result coincides with the one in Refs. \cite{Cai,Boudjedaa,Chetouani2,bayrak2007}.

%---------------------------------------------------------------------------------------
\subsubsection{Screened potential}
%---------------------------------------------------------------------------------------

By choosing $q=1,2\alpha =\eta ,a=\frac{\hbar^2}{2m}\eta^2 \sigma\left( \sigma+1\right) ,b=\frac{2m\eta
	Ze^{2}}{\hbar ^{2}}-a,$ and $l=0$ the potential (\ref{a.1}) is reduced to the screened
potential \cite{Sadoun2}. Here $\sigma$ represents an adjustment parameter linked to the screening distance.
The parameters in Eq. \eqref{a.31} reduces to:
\begin{equation}
	N_{\sigma}=\frac{\eta}{2} \left( \sigma+1\right) 
\end{equation}%

The discrete energy spectrum of the screened potential can be deduced from equation (\ref{a.30})%
\begin{equation}
	E_{N}=-\frac{\eta ^{2}\hbar ^{2}}{8m}\left( \frac{2mZe^{2}}{\eta \hbar ^{2}N}%
	-N\right) ^{2},
\end{equation}
with $N=n+\sigma+1$.

	%---------------------------------------------------------------------------------------
	\subsubsection{Linear combination of Hulth\'{e}n and Yukawa potentials}
	%---------------------------------------------------------------------------------------

	A linear combination of the deformed Hulth\'{e}n potential and the Yukawa potential is obtained by taking $a=0,c=0,q=1,2\alpha =\eta ,b=\eta Ze^{2}$ et $l=0$;
	\begin{equation}
		V\left( r\right) =-\frac{\eta Ze^{2}}{e^{\eta r}-q}-\frac{V_{0}e^{-\eta r}}{r}.
		\label{a.50}
	\end{equation}
	
	With these changes, the parameters (\ref{a.31}) and (\ref{a.33}) are given by%
\begin{equation}
	E_{n,0}
	=
	-\frac{\hbar^2}{2m}
	\left[
	\frac{m\left(Ze^2+\eta V_0\right)}{\hbar^2\,(n+1)}
	-\frac{\eta}{2}(n+1)
	\right]^2 .
\end{equation}
which corresponds to the same expression of the energy spectrum obtained by asymptotic iteration method by taken $l=0$ \cite{adebimpe1}.

	%---------------------------------------------------------------------------------------%---------------------------------------------------------------------------------------

%---------------------------------------------------------------------------------------
\section{Thermodynamic properties}\label{Sec:TP}
%---------------------------------------------------------------------------------------

In molecular physics, the thermodynamic properties of a molecule can be obtained from the vibrational partition function, which in discrete form is written as the sum over all possible vibrational energy levels \cite{vibpartitionf1,vibpartitionf2}:
\begin{equation}
	Z_{vib}(\beta)=\sum_{n=0}^{\lambda}e^{-\beta E_{n}},\quad \beta=\frac{1}{k_BT},
\end{equation}
where $\lambda_{max}$, $k_B$, and $T$ denote the maximum vibrational quantum number, the Boltzmann constant, and the temperature, respectively. The maximum vibrational quantum number $\lambda_{max}$ is obtained by solving the following condition \cite{thermodynamic2023}:
\begin{equation}
	\frac{\partial E^{q}_{n,l}}{\partial n}|_{n=\lambda_{max}}=0,
\end{equation}
where the solution of this equation for different molecules is presented in the next section.

The energy spectrum given in Eq. \eqref{a.30} can be simplified as follows:
\begin{equation}
	E^q_{n,l}=-\Lambda\left(\frac{\eta_1}{n+\eta_2}-(n+\eta_2)\right)^2,
\end{equation}
where:
\begin{align}
	\Lambda&=\frac{\alpha \hbar^2}{2m}, \quad\eta_1=\frac{m}{2\hbar ^{2}}\left( \frac{b+2\alpha c}{\alpha ^{2}q}+\frac{a}{\alpha^2 q^{2}}\right),\quad	\eta_2=\frac{1}{2}\left( 1+\sqrt{1+\frac{4l\left( l+1\right) }{q}+\frac{2m}{\hbar^{2}}
		\left( \frac{a}{\alpha^2 q^{2}}\right)}\right).
\end{align}
The partition function is then written as:
\begin{equation}
	Z_{vib}(\beta)=\sum_{n=0}^{\lambda}e^{\beta \Lambda\left(\frac{\eta_1}{n+\eta_2}-(n+\eta_2)\right)^2},
\end{equation}
To compute the vibrational partition function for the finite summation $\lambda_{\text{max}}$, we use the Poisson summation formula, which is given by \cite{vibpartitionf1,vibpartitionf2}:
\begin{equation}
	\sum_{n=0}^{\lambda_{\text{max}}} f(n) = \frac{1}{2}\left[f(0) - f(\lambda_{\text{max}}+ 1)\right] + \sum_{m=-\infty}^{\infty} \int_{0}^{\lambda_{\text{max}} + 1} f(x) e^{-i2\pi mx} dx
\end{equation}
and, at leading order of approximation, corresponding to $m=0$ and the semiclassical case \cite{strekalov1}, we can write:
\begin{equation}
	\sum_{n=0}^{\lambda_{\text{max}}} f(n) = \frac{1}{2}\left[f(0) - f(\lambda_{\text{max}} + 1)\right] + \int_{0}^{\lambda_{\text{max}} + 1} f(x) dx
\end{equation}
In our case, the function $f(n)$ is given by:
\begin{equation}
	f(n)=e^{\beta \Lambda\left(\frac{\eta_1}{n+\eta_2}-(n+\eta_2)\right)^2},
\end{equation}
Then the vibrational partition function becomes:
\begin{equation}\label{eq:vpf3}
	Z_{vib}(\beta) = \frac{1}{2}\left[e^{\beta \Lambda \rho_{1}^{2}} - e^{\beta \Lambda \rho_{2}^{2}}\right] + \int_{0}^{\lambda_{\text{max}} + 1} e^{\beta\lambda\left(\frac{\eta_{1}}{x + \eta_{2}} - (x + \eta_{2})\right)^{2}} dx
\end{equation}
where:
\begin{equation}
	\rho_{1}=\frac{\eta_{1}}{\eta_{2}} -  \eta_{2},\,\, \rho_{2}=\frac{\eta_{1}}{\lambda_{\text{max}} + 1 + \eta_{2}} - (\lambda_{\text{max}} + 1 + \eta_{2}).
\end{equation}
To compute the integral in the last term of Eq. \eqref{eq:vpf3}, we introduce a new variable,
\[
y=\frac{\eta_{1}}{x + \eta_{2}} - (x + \eta_{2}).
\]
The integral can then be evaluated as follows:
\begin{equation}
	\begin{split}
		\int_{0}^{\lambda_{\text{max}} + 1} e^{\beta \Lambda\left(\frac{\eta_{1}}{x + \eta_{2}} - x + \eta_{2}\right)^{2}} dx &= \frac{1}{2}\int_{\rho_{1}}^{\rho_{2}} e^{\beta\Lambda y^{2}} \left(\frac{y}{\sqrt{y^{2} + 4\eta_{1}}} - 1\right) dy \\
		&= \frac{1}{2} \sqrt{\frac{\pi}{\beta\Lambda}} \Bigg[\operatorname{erfi}\left(\sqrt{\beta\Lambda} \rho_{1}\right) - \operatorname{erfi}\left(\sqrt{\beta\Lambda} \rho_{2}\right)  \\
		&\quad  + e^{-4\beta\Lambda\eta_{1}}\bigg(- \operatorname{erfi}\left(\sqrt{\beta\Lambda(4\eta_{1} + \rho_{1}^2)}\right) + \operatorname{erfi}\left(\sqrt{\beta\Lambda(4\eta_{1} + \rho_{2}^2)}\right)\bigg) \Bigg]
	\end{split}
\end{equation}
where $\operatorname{erfi}(x)$ is the imaginary error function\footnote{The complex error function is defined by: $\operatorname{erfi}(z) = -i \operatorname{erf}(iz) = \frac{2}{\sqrt{\pi}} \int_{0}^{z} e^{t^{2}} dt$}. The final vibrational partition function is written as:
\begin{align}\label{eq:Zvib}
	Z_{vib}(\beta,\lambda_{\text{max}}) &= \frac{1}{2}\left[e^{\beta \Lambda \rho_{1}^{2}} - e^{\beta \Lambda \rho_{2}^{2}}\right]+\frac{1}{4} \sqrt{\frac{\pi}{\beta\Lambda}} \Bigg[\operatorname{erfi}\left(\sqrt{\beta\Lambda} \rho_{1}\right) - \operatorname{erfi}\left(\sqrt{\beta\Lambda} \rho_{2}\right)  \notag\\
	&\quad  + e^{-4\beta\Lambda\eta_{1}}\bigg(- \operatorname{erfi}\left(\sqrt{\beta\Lambda(4\eta_{1} + \rho_{1}^2)}\right) + \operatorname{erfi}\left(\sqrt{\beta\Lambda(4\eta_{1} + \rho_{2}^2)}\right)\bigg) \Bigg].
\end{align}

Using the above vibrational partition function, one can obtain all thermodynamic properties of a diatomic molecule described by the Yukawa plus four-parameter diatomic potential.

\subsubsection{Vibrational free energy}

The free energy of the molecule is expressed as follows:
\begin{align}
	F(\beta)&=-k_BT\ln Z_{vib}(\beta),\notag\\
	&=-\frac{1}{\beta}\ln\Bigg[\frac{1}{2}\left[e^{\beta \Lambda \rho_{1}^{2}} - e^{\beta \Lambda \rho_{2}^{2}}\right]+\frac{1}{4} \sqrt{\frac{\pi}{\beta\Lambda}} \Bigg[\operatorname{erfi}\left(\sqrt{\beta\Lambda} \rho_{1}\right) - \operatorname{erfi}\left(\sqrt{\beta\Lambda} \rho_{2}\right)  \notag\\
	&\quad  + e^{-4\beta\Lambda\eta_{1}}\bigg(- \operatorname{erfi}\left(\sqrt{\beta\Lambda(4\eta_{1} + \rho_{1}^2)}\right) + \operatorname{erfi}\left(\sqrt{\beta\Lambda(4\eta_{1} + \rho_{2}^2)}\right)\bigg) \Bigg]\Bigg].\label{eq:Fvib}
\end{align}

\subsubsection{Vibrational mean energy}
By definition, the vibrational mean energy can be evaluated using the above partition function in Eq. \eqref{eq:Zvib}:
\begin{align}
	U_{vib}(\beta)&=-\frac{\partial \ln Z_{vib}}{\partial \beta},\notag\\
	&=\frac{4(\beta\Lambda)^{3/2} e^{4 \beta \eta_1 \Lambda }  e^{\beta  \Lambda  y^2} y^2 \Big|^{\rho_2}_{\rho_1}+\sqrt{\pi} \left(\operatorname{erfi}\left(\sqrt{\beta  \Lambda } \sqrt{4 \eta _1+y^2}\right)\Big|^{\rho_2}_{\rho_1}-e^{4 \beta \Lambda  \eta _1}\operatorname{erfi}\left(y \sqrt{\beta  \Lambda }\right)\Big|^{\rho_2}_{\rho_1}\right)}{2 \beta\left(\sqrt{\pi}\left(\operatorname{erfi}\left(\sqrt{\beta  \Lambda }
		\sqrt{4 \eta_1+y^2}\right)\Big|^{\rho_2}_{\rho_1}-e^{4 \beta \eta_1 \Lambda } \operatorname{erfi}\left(y \sqrt{\beta  \Lambda }\right)\Big|^{\rho_2}_{\rho_1}\right)-2e^{4 \beta \eta_1 \Lambda}e^{\beta  \Lambda  y^2}\Big|^{\rho_2}_{\rho_1}\right)}\notag\\
	&+\frac{ 2\sqrt{\beta\Lambda} e^{4 \beta \eta_1 \Lambda }\left(y -\sqrt{4 \eta _1+y^2} \right)e^{\beta  \Lambda  y^2}\Big|^{\rho_2}_{\rho_1}+8 \sqrt{\pi} \beta  \Lambda\eta _1    e^{-4 \beta  \eta _1 \Lambda } \operatorname{erfi}\left(\sqrt{\beta  \Lambda } \sqrt{4
			\eta _1+y^2}\right)\Big|^{\rho_2}_{\rho_1}}{2 \beta 
		\Bigg(\sqrt{\pi} \left( \operatorname{erfi}\left(\sqrt{\beta  \Lambda }
		\sqrt{4 \eta_1+y^2}\right)\Big|^{\rho_2}_{\rho_1}-e^{4 \beta  \Lambda \eta_1 }\operatorname{erfi}\left(y \sqrt{\beta  \Lambda }\right)\Big|^{\rho_2}_{\rho_1}\right)-2e^{4 \beta \Lambda \eta_1}e^{\beta  \Lambda  y^2}\Big|^{\rho_2}_{\rho_1}\Bigg)}\label{eq:Uvib}
\end{align}

\subsubsection{Vibrational heat capacity}

The heat capacity of this system is defined as the derivative of the mean energy with respect to $\beta$ and is given by:
\begin{align}
	C_{vib}(\beta)&=-k_B\beta^2 \frac{\partial U_{vib}(\beta)}{\partial \beta},
\end{align}
By using the vibrational mean energy expression in Eq. \eqref{eq:Uvib}, we find:
\begin{align}
	C_{vib}(\beta)&=\frac{-2 (\beta  \Lambda )^{3/2} e^{2 \beta  \Lambda  \left(4 \eta _1+y^2\right)}\Gamma_1(y)\Big|^{\rho_2}_{\rho_1}+\pi  \sqrt{\beta  \Lambda } e^{8 \beta  \eta _1 \Lambda } \operatorname{erfi}\left(y \sqrt{\beta  \Lambda }\right)^2\Big|^{\rho_2}_{\rho_1}+\pi  \sqrt{\beta  \Lambda } \operatorname{erfi}\left(\sqrt{\beta  \Lambda } \sqrt{4 \eta _1+y^2}\right)^2\Big|^{\rho_2}_{\rho_1}}{2  \left(-\sqrt{\pi } \operatorname{erfi}\left(\sqrt{\beta\Lambda}\sqrt{4\eta_1+y^2}\right)\Big|^{\rho_2}_{\rho_1}+\sqrt{\pi } e^{4\beta  \eta _1 \Lambda } \operatorname{erfi}\left(y \sqrt{\beta\Lambda }\right)\Big|^{\rho_2}_{\rho_1}+2 \sqrt{\beta  \Lambda } e^{\beta\Lambda\left(4 \eta_1+y^2\right)}\Big|^{\rho_2}_{\rho_1}\right)^2}\notag\\
	&+\frac{-\sqrt{\pi }\beta\Lambda e^{\beta\Lambda\left(4\eta_1+y^2\right)}\Gamma_2(y)\operatorname{erfi}\left(\sqrt{\beta\Lambda } \sqrt{4 \eta _1+y^2}\right)\Big|^{\rho_2}_{\rho_1}+\sqrt{\pi } \operatorname{erfi}\left(y \sqrt{\beta  \Lambda }\right)\Big(\sqrt{\beta  \Lambda }e^{\beta  \Lambda  \left(8 \eta _1+y^2\right)}\Gamma_4(y)-\Gamma_3(y)\Big)\Big|^{\rho_2}_{\rho_1}}{2 \left(-\sqrt{\pi } \operatorname{erfi}\left(\sqrt{\beta\Lambda}\sqrt{4\eta_1+y^2}\right)\Big|^{\rho_2}_{\rho_1}+\sqrt{\pi } e^{4\beta  \eta _1 \Lambda } \operatorname{erfi}\left(y \sqrt{\beta\Lambda }\right)\Big|^{\rho_2}_{\rho_1}+2 \sqrt{\beta  \Lambda } e^{\beta\Lambda\left(4 \eta_1+y^2\right)}\Big|^{\rho_2}_{\rho_1}\right)^2}.\label{eq:Cvib}
\end{align}
where
\begin{align}
	\Gamma_1(y)&=\left(4 \eta _1+2 \beta  \Lambda  y^3+2 y^2 \left(1-\beta  \Lambda 
	\sqrt{4 \eta _1+y^2}\right)-8 \beta  \eta _1 \Lambda  \sqrt{4 \eta _1+y^2}+y \left(3-2 \sqrt{4 \eta _1+y^2}\right)-3 \sqrt{4 \eta
		_1+y^2}\right),\\
	\Gamma_2(y)&=\Big(64 \beta ^2 \eta _1^2 \Lambda ^2+4 \beta ^2 \Lambda ^2 y^4+2
	\beta  \Lambda  y^3+2 \beta  \Lambda  y^2 \left(16 \beta  \eta _1 \Lambda -\sqrt{4 \eta _1+y^2}+2\right)-8 \beta  \eta _1 \Lambda 
	\left(\sqrt{4 \eta _1+y^2}-2\right)\notag\\
	&+\sqrt{4 \eta _1+y^2}+y \left(16 \beta  \eta _1 \Lambda -1\right)+3\Big),\\
	\Gamma_3(y)	&=2 \sqrt{\pi }  e^{4 \beta  \eta _1 \Lambda } \left(16 \beta ^2 \eta _1^2 \Lambda ^2+1\right)\text{erfi}\left(\sqrt{\beta  \Lambda } \sqrt{4 \eta _1+y^2}\right),\\
	\Gamma_4(y)	&=4 \beta ^2 \Lambda ^2 y^4+2 \beta  \Lambda  y^3-2 \beta  \Lambda  y^2 \left(\sqrt{4 \eta _1+y^2}-2\right)+8 \beta  \eta _1 \Lambda  \sqrt{4
		\eta _1+y^2}+\sqrt{4 \eta _1+y^2}-y+3
\end{align}

\subsubsection{Vibrational entropy}
The entropy of this system is given by the following definition:
\begin{align}
	S(\beta)&=k_B\ln Z_{vib}-k_B\beta\frac{\partial\ln Z_{vib}}{\partial \beta}.
\end{align}
Using the explicit expression of the vibrational partition function in Eq. \eqref{eq:Zvib}, we obtain the following expression for the vibrational entropy:
\begin{align}
	S(\beta)&=k_B\ln\Bigg[-\frac{1}{2} e^{\beta \Lambda \rho_{2}^{2}}\Big|^{\rho_2}_{\rho_1}+\frac{1}{4} \sqrt{\frac{\pi}{\beta\Lambda}} \Bigg[- \operatorname{erfi}\left(\sqrt{\beta\Lambda} \rho_{2}\right)\Big|^{\rho_2}_{\rho_1} + e^{-4\beta\Lambda\eta_{1}} \operatorname{erfi}\left(\sqrt{\beta\Lambda(4\eta_{1} + \rho_{2}^2)}\right)\Big|^{\rho_2}_{\rho_1} \Bigg]\Bigg]\notag\\
	&+k_B\beta\Bigg[\frac{\Lambda  y^2 e^{\beta  \Lambda  y^2}\Big|^{\rho_2}_{\rho_1}+\frac{1}{4}\sqrt{\frac{\pi}{\beta ^3 \Lambda }} \left(e^{-4 \beta  \eta _1
			\Lambda } \operatorname{erfi}\left(\sqrt{\beta  \Lambda } \sqrt{4 \eta _1+y^2}\right)\Big|^{\rho_2}_{\rho_1}-\operatorname{erfi}\left(y \sqrt{\beta  \Lambda }\right)\Big|^{\rho_2}_{\rho_1}\right)}{\frac{1}{2}\sqrt{\frac{\pi }{\beta  \Lambda }} \left(e^{-4 \beta \eta_1 \Lambda } \operatorname{erfi}\left(\sqrt{\beta  \Lambda }
		\sqrt{4 \eta_1+y^2}\right)\Big|^{\rho_2}_{\rho_1}-\operatorname{erfi}\left(y \sqrt{\beta  \Lambda }\right)\Big|^{\rho_2}_{\rho_1}\right)-e^{\beta  \Lambda  y^2}\Big|^{\rho_2}_{\rho_1}}\notag\\
	&-\frac{\sqrt{4 \eta _1+y^2} e^{\beta  \Lambda  y^2}\Big|^{\rho_2}_{\rho_1}-  y e^{\beta  \Lambda  y^2}\Big|^{\rho_2}_{\rho_1}-4 \sqrt{\pi \beta  \Lambda} \eta _1    e^{-4 \beta  \eta _1 \Lambda } \operatorname{erfi}\left(\sqrt{\beta  \Lambda } \sqrt{4
			\eta _1+y^2}\right)\Big|^{\rho_2}_{\rho_1}}{2 \beta 
		\Bigg(\frac{1}{2}\sqrt{\frac{\pi }{\beta  \Lambda }} \left(e^{-4 \beta \eta_1 \Lambda } \operatorname{erfi}\left(\sqrt{\beta  \Lambda }
		\sqrt{4 \eta_1+y^2}\right)\Big|^{\rho_2}_{\rho_1}-\operatorname{erfi}\left(y \sqrt{\beta  \Lambda }\right)\Big|^{\rho_2}_{\rho_1}\right)-e^{\beta  \Lambda  y^2}\Big|^{\rho_2}_{\rho_1}\Bigg)}\Bigg].\label{eq:Svib}
\end{align}
	%---------------------------------------------------------------------------------------
	%---------------------------------------------------------------------------------------
\section{Results and Discussion}\label{Sec:RD}

In this section, we present our main results and discussion, together with several plots of the thermodynamic properties discussed above for a few diatomic molecules. 

\begin{figure}[ht]
	\centering
	\includegraphics[width=0.45\textwidth]{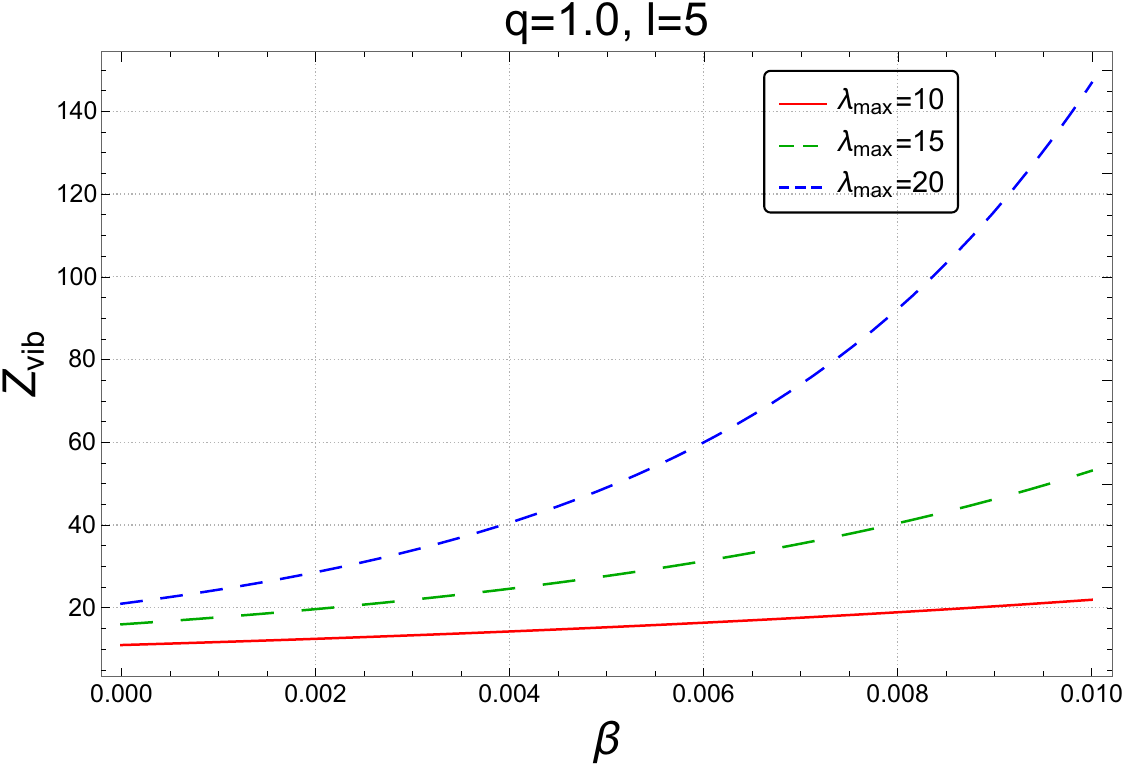}
	\includegraphics[width=0.45\textwidth]{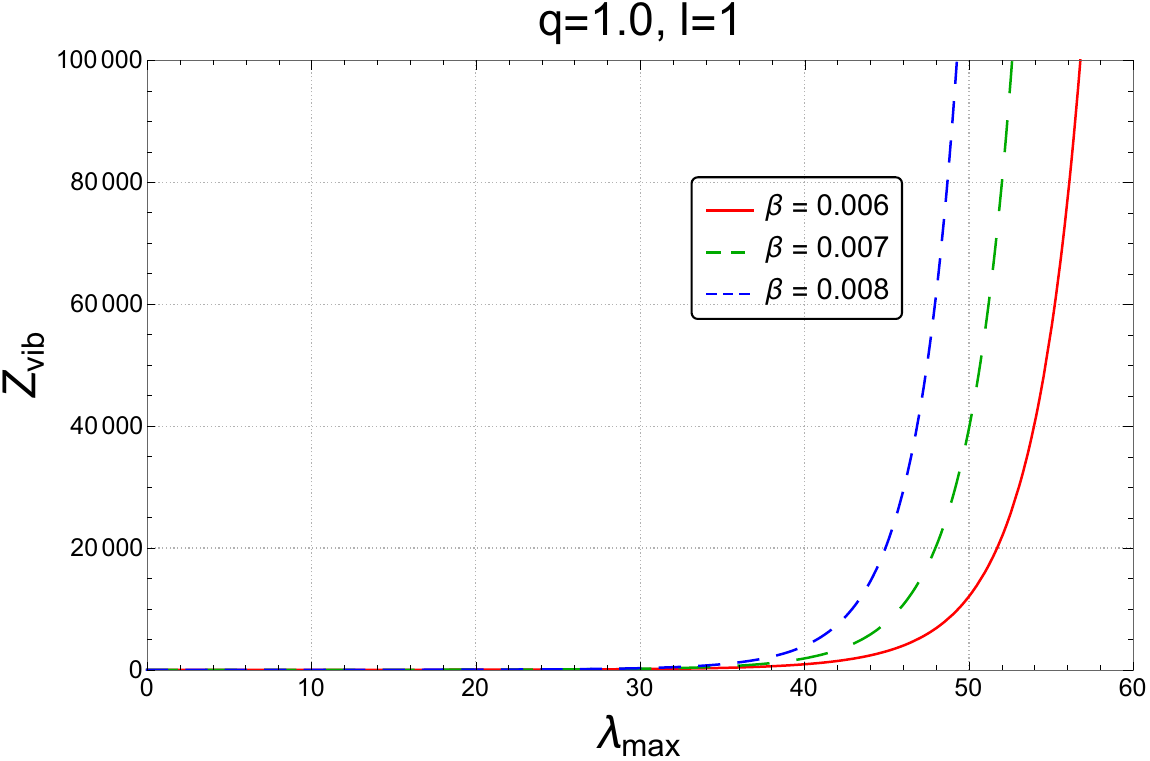}
	\includegraphics[width=0.45\textwidth]{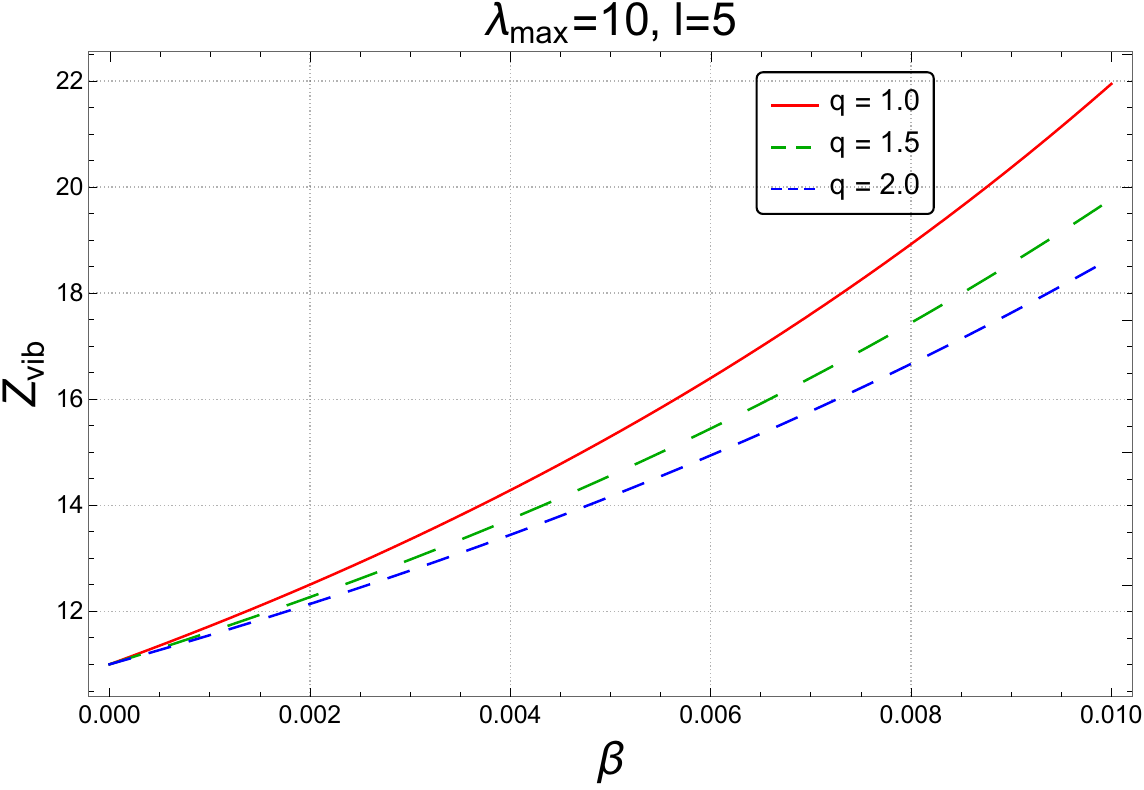}
	\includegraphics[width=0.45\textwidth]{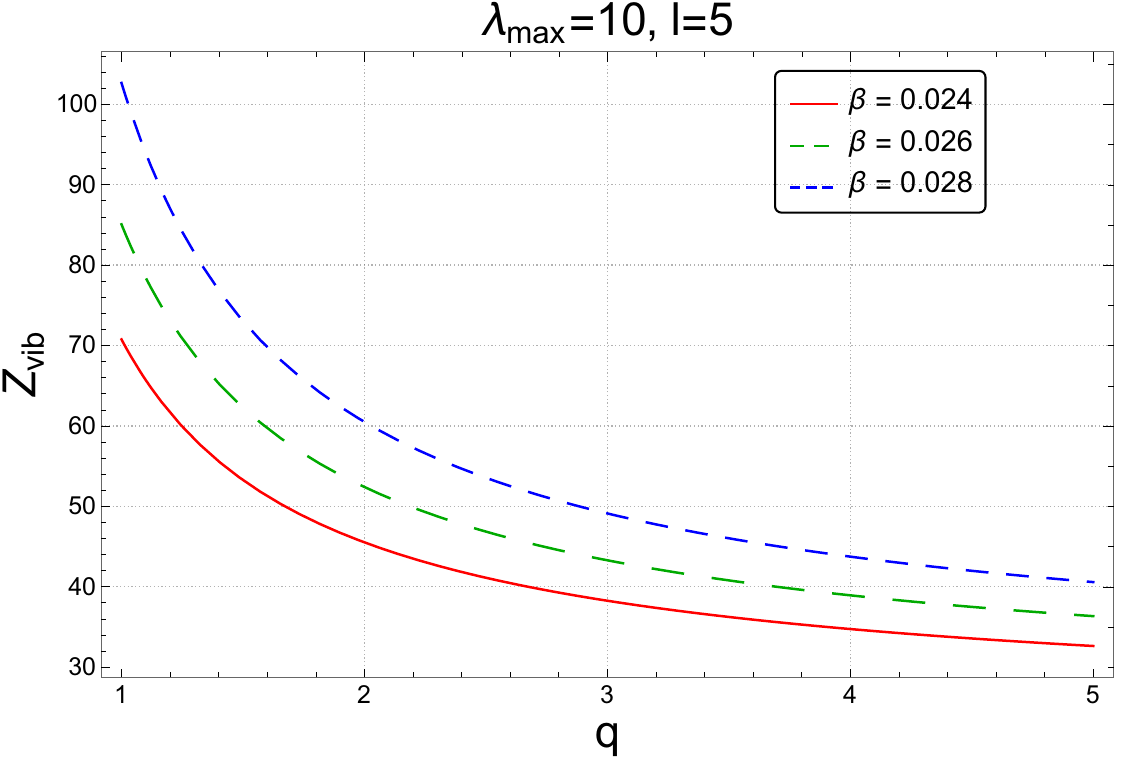}
	\includegraphics[width=0.45\textwidth]{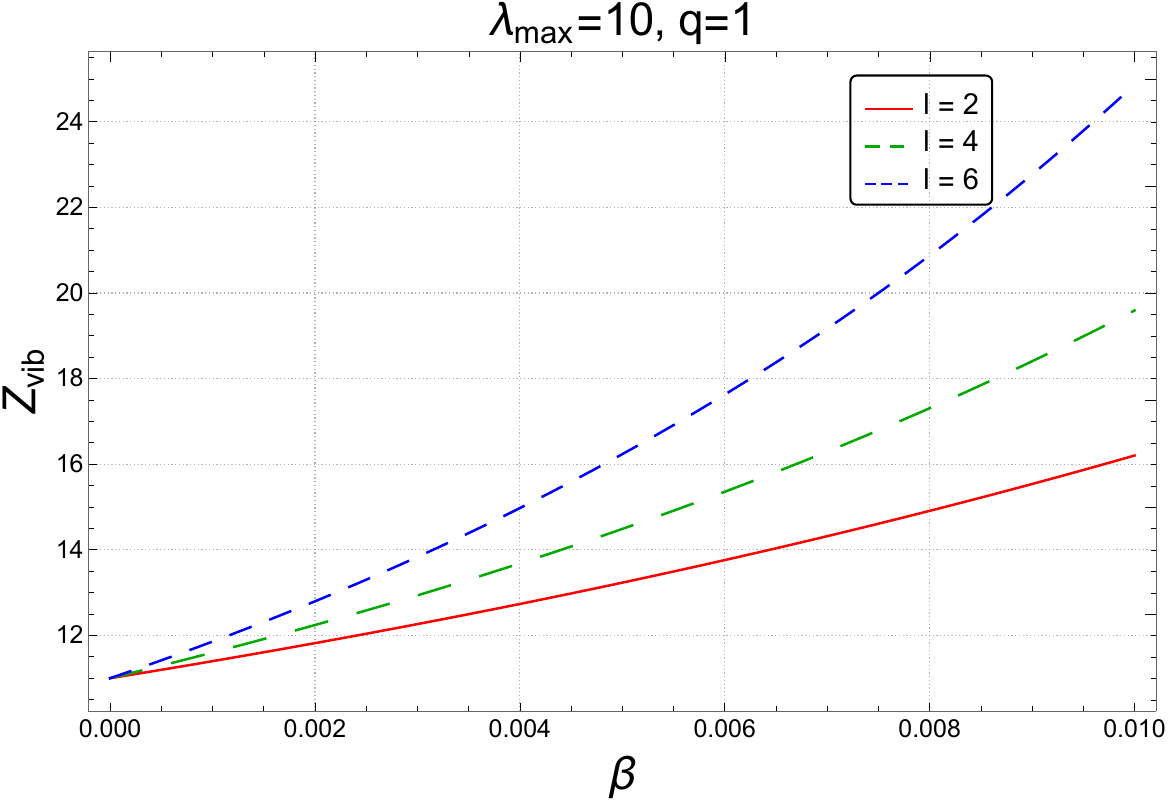}
	\includegraphics[width=0.45\textwidth]{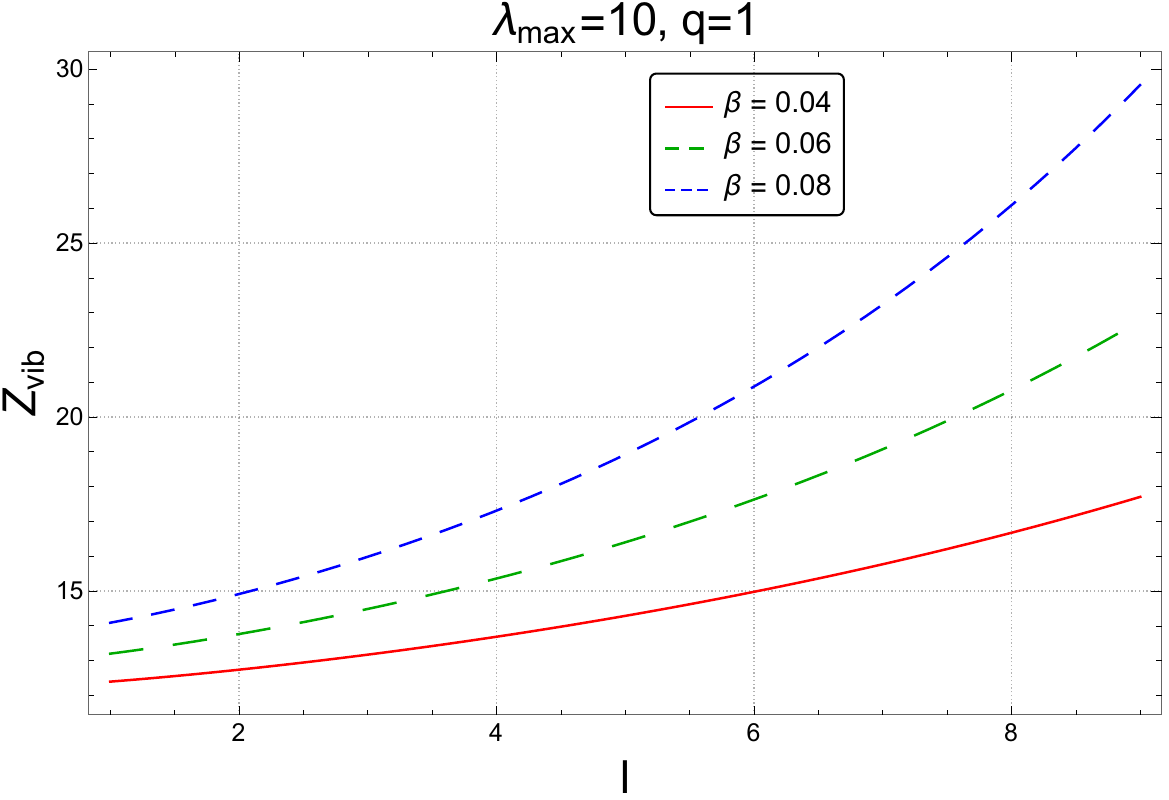}
	\caption{Partition function as a function of $\beta$, the deformation parameter $q$, and the quantum number $l$.}  \label{fig:z}
\end{figure}

In Fig.~\ref{fig:z}, we illustrate the behavior of the vibrational partition function for our system as a function of $\beta$, the deformation parameter $q$, and the orbital quantum number $l$. As shown, the vibrational partition function as a function of $\beta$ and the quantum number $\lambda_{\text{max}}$ exhibits the same behavior: it increases with $\beta$ and grows exponentially with $\lambda_{\text{max}}$ (first row). For the deformation parameter $q$ (middle row), the behavior is opposite: the vibrational partition function decreases as $q$ increases. In the last row, we show the effect of the orbital quantum number $l$ on the vibrational partition function, which is similar to the effect of $\beta$: the vibrational partition function increases with $l$.

\subsection{Application to diatomic molecules}

In what follows, we summarize some diatomic molecules and their spectroscopic parameters in the following table:
\begin{table}[h]
	\begin{center}
		\caption{Some selected diatomic molecules with their spectroscopic parameters \cite{spectroscopic1}.}\label{tab5}
		\begin{tabular}{c c c c c c}
			\hline
			\hline
			Molecule& $\lambda_{max}$ (this work)& $r_e(\dot{A})$ & $D_e$ ($eV$) &  $m(amu)$ & $\alpha$ ($\dot{A}^{-1}$)\\
			\hline
			\hline
			$H_2$ &	8 & 0.7416 & 4.74460 & 0.503910 & 1.61890\\		
			\hline
			$I_2$ & 58 & 2.6620  & 1.55560 & 63.452235 & 1.86430\\
			\hline
			$LiH$ & 13 & 1.5956 & 2.51527 & 0.880122 & 1.12800\\
			\hline
			$CO$ & 40 & 1.1283 & 11.2256 & 6.860672 & 2.29940\\
			\hline
			$HCL$ & 11 & 1.2746 & 4.61903 & 0.980105 & 1.86770\\
			\hline
			$NO$ & 38 & 1.1508 & 8.04373 & 7.468441 & 2.75340 \\
			\hline
			\hline
		\end{tabular}
	\end{center}
\end{table}

As a first application, we compute the numerical energy for a few molecules, as summarized in the following table:
\begin{table}[h]
	\begin{center}
		\caption{Numerical values of the energy (in eV) for a few molecules for fixed parameters $l=0$ and different $q$ and $n$.}\label{tab4}
		\begin{tabular}{c| c|c c c c c c}
			\hline
			\hline
			Molecule & Energy (eV) & $q=1.0$ & $q=1.2$ & $q=1.4$ & $q=1.6$ & $q=1.8$ & $q=2.0$ \\
			\hline
			\hline
			&  & -4.1375 & -4.1018 & -4.0664 & -4.0313 & -3.9965 & -3.962 \\
			
			&  & -3.06476 & -2.98707 & -2.91202 & -2.83949 & -2.76938 & -2.70157 \\
			
			$H_2$& $E_{n=0,4}^{q\geq1}$	& -2.19217 & -2.09759 & -2.00832 & -1.92399 & -1.84425 & -1.76879 \\
			
			&  & -1.49449 & -1.39976 & -1.31241 & -1.23174 & -1.1571 & -1.08796 \\
			
			&  & -0.950793 & -0.866656 & -0.791001 & -0.72281 & -0.661209 & -0.605443 \\
			\hline
			& & -1.57224 & -1.57217 & -1.5721 & -1.57204 & -1.57197 & -1.5719 \\
			& & -1.52587 & -1.52567 & -1.52548 & -1.52528 & -1.52509 & -1.52489 \\
			$I_2$	& $E_{n=0,4}^{q\geq1}$ & -1.4802 & -1.47989 & -1.47957 & -1.47925 & -1.47893 & -1.47862 \\
			& & -1.43524 & -1.43481 & -1.43437 & -1.43394 & -1.43351 & -1.43307 \\
			& & -1.39098 & -1.39044 & -1.38989 & -1.38935 & -1.3888 & -1.38826 \\
			\hline
			&	& -2.32299 & -2.31639 & -2.3098 & -2.30324 & -2.29669 & -2.29017 \\
			
			&	& -1.94441 & -1.92774 & -1.91127 & -1.89499 & -1.8789 & -1.863 \\
			
			$LiH$ & $E_{n=0,4}^{q\geq1}$ & -1.60451 & -1.581 & -1.55798 & -1.53541 & -1.51328 & -1.4916 \\
			
			&	  & -1.30158 & -1.27396 & -1.24713 & -1.22105 & -1.19569 & -1.17103 \\
			
			&	& -1.03404 & -1.00457 & -0.976175 & -0.94881 & -0.922425 & -0.896977 \\
			\hline
			& & -10.9743 & -10.97 & -10.9657 & -10.9614 & -10.9571 & -10.9528 \\
			& & -10.4081 & -10.3959 & -10.3836 & -10.3713 & -10.3591 & -10.3469 \\
			$CO$ & $E_{n=0,4}^{q\geq1}$ & -9.85812 & -9.83854 & -9.81902 & -9.79956 & -9.78014 & -9.76079 \\
			& & -9.32412 & -9.29792 & -9.27182 & -9.24583 & -9.21995 & -9.19417 \\
			& & -8.80603 & -8.77385 & -8.74184 & -8.71 & -8.67832 & -8.64681 \\
			\hline
			&	& -4.26735 & -4.26009 & -4.25283 & -4.24559 & -4.23836 & -4.23114 \\
			
			&	& -3.52411 & -3.50575 & -3.48752 & -3.46943 & -3.45146 & -3.43362  \\
			
			$HCL$ & $E_{n=0,4}^{q\geq1}$ & -2.85746 & -2.83165 & -2.80616 & -2.78098 & -2.7561 & -2.73153 \\
			
			&	 & -2.26547 & -2.23533 & -2.20571 & -2.17662 & -2.14802 & -2.11992 \\
			
			&  & -1.74624 & -1.71441 & -1.68332 & -1.65294 & -1.62324 & -1.59421 \\
			\hline
			& & -7.79932 & -7.79706 & -7.7948 & -7.79254 & -7.79028 & -7.78802 \\
			& & -7.27216 & -7.26577 & -7.2594 & -7.25303 & -7.24667 & -7.24032 \\
			$NO$ & $E_{n=0,4}^{q\geq1}$ & -6.76418 & -6.75413 & -6.74411 & -6.7341 & -6.72412 & -6.71416 \\
			& & -6.2753 & -6.26203 & -6.2488 & -6.23561 & -6.22246 & -6.20936 \\
			& & -5.80543 & -5.78936 & -5.77336 & -5.75741 & -5.74153 & -5.72571 \\
			\hline
			\hline
		\end{tabular}
	\end{center}
\end{table}

In Table~\ref{tab4}, all molecules exhibit the same qualitative behavior across rows and columns. The energy decreases as the energy level $n$ increases for a fixed deformation parameter $q$ (along each column), and it also decreases as $q$ increases for a fixed $n$ (along each row). This result confirms the consistency of the model for these molecules. We observe that the molecules $CO$, $NO$, and $LiH$ show a slow variation of the energy along both rows and columns. The energy difference $\Delta E_{n}^{q}$ is larger when $n$ changes than when $q$ varies, so the deformation parameter has no major impact for these molecules. Note that $CO$ and $NO$ have close spectroscopic parameters. The molecules $H_2$ and $HCl$ show a faster variation of $\Delta E_{n}^{q}$, indicating that the deformation parameter has a stronger influence on the energy for a given energy level $n$. The heavier molecule $I_2$ exhibits the slowest variation: its energy levels are very close, and the influence of $q$ is negligible. 

For more details on the energy spectrum of the model, the general effects of the screening parameter and the deformation parameter $q$ on the energy spectrum, for different quantum numbers $n$ and $l$, are shown in Figs.~\ref{fig01} and \ref{fig02}, respectively. The influence of the dissociation energy $D_e$ and the depth of the potential well $V_0$ is shown in Fig.~\ref{fig03}. \\  
For these different molecules, we investigate the behavior of the partition function: 
\begin{figure}[ht]
	\centering
	\includegraphics[width=0.33\textwidth]{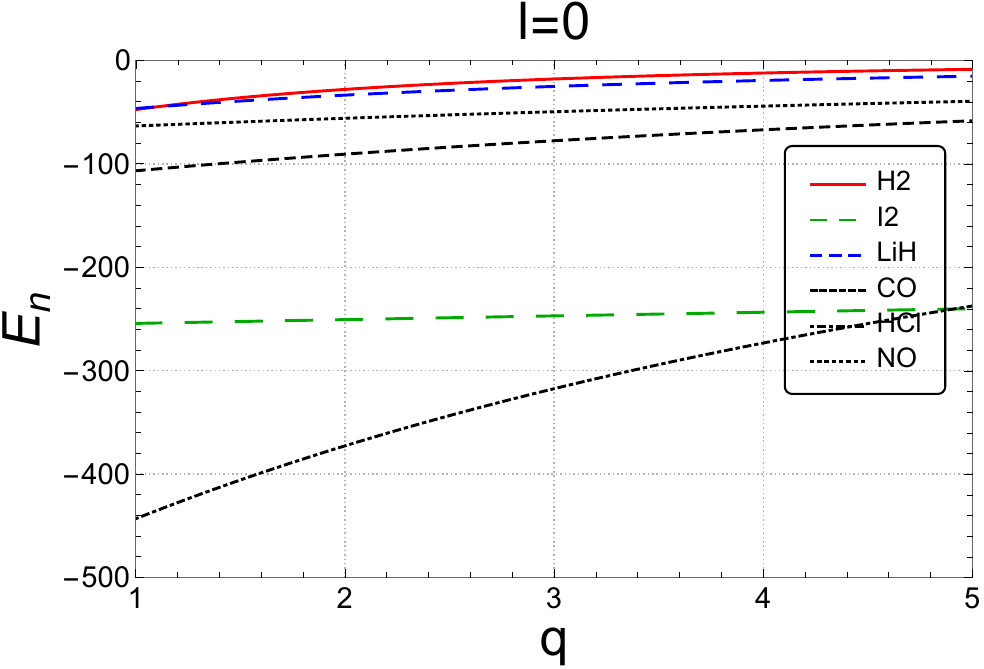}
	\includegraphics[width=0.32\textwidth]{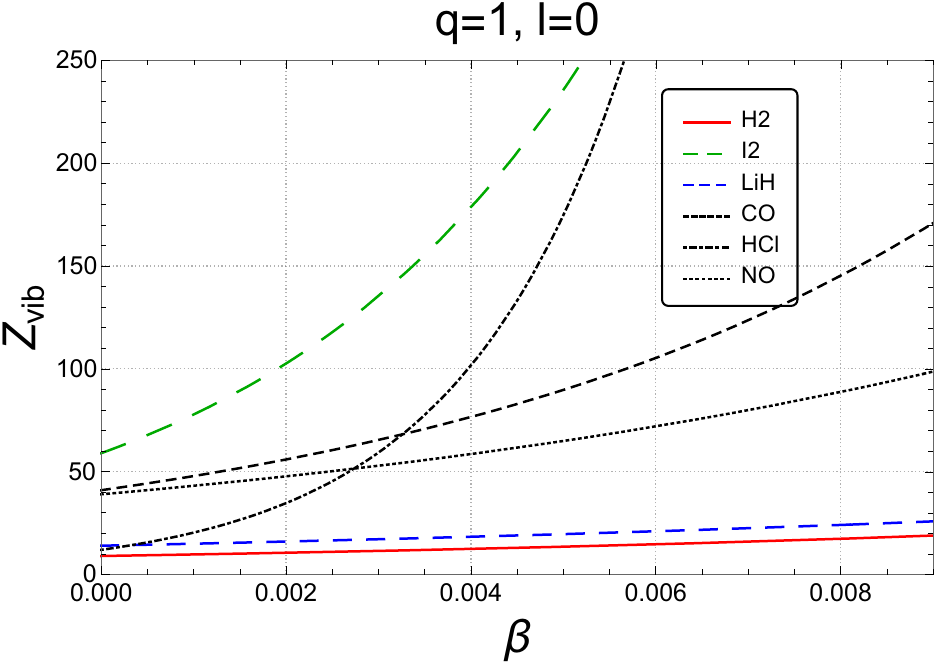}
	\includegraphics[width=0.335\textwidth]{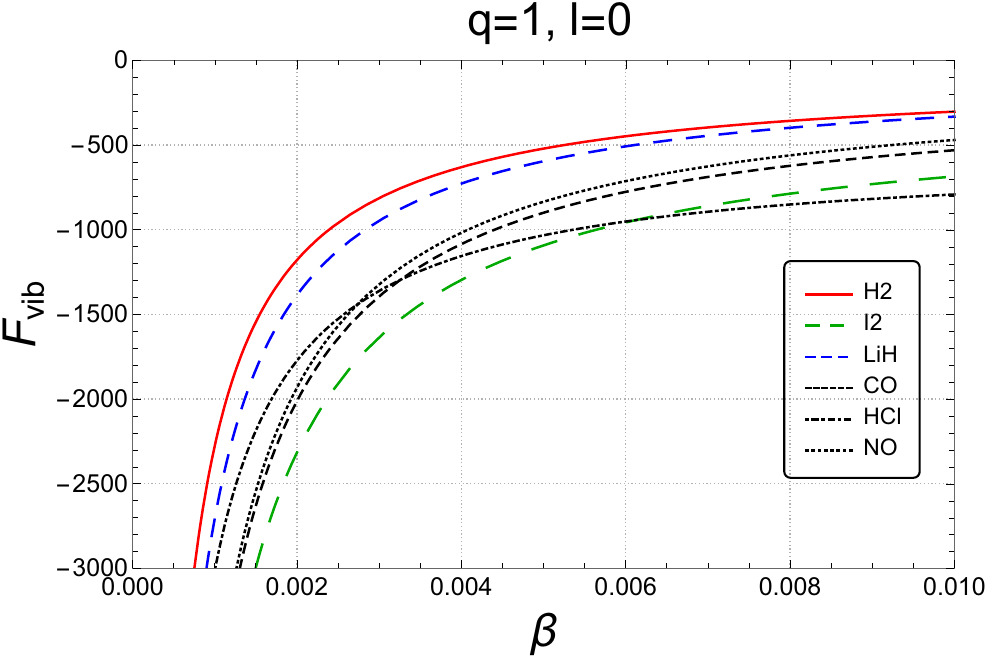}
	\includegraphics[width=0.325\textwidth]{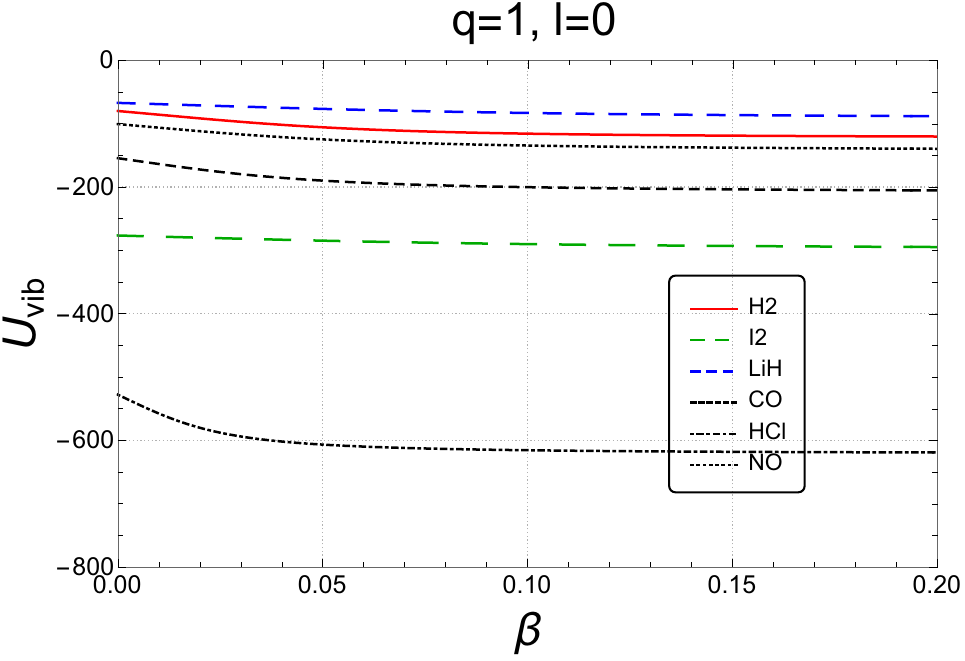}
	\includegraphics[width=0.33\textwidth]{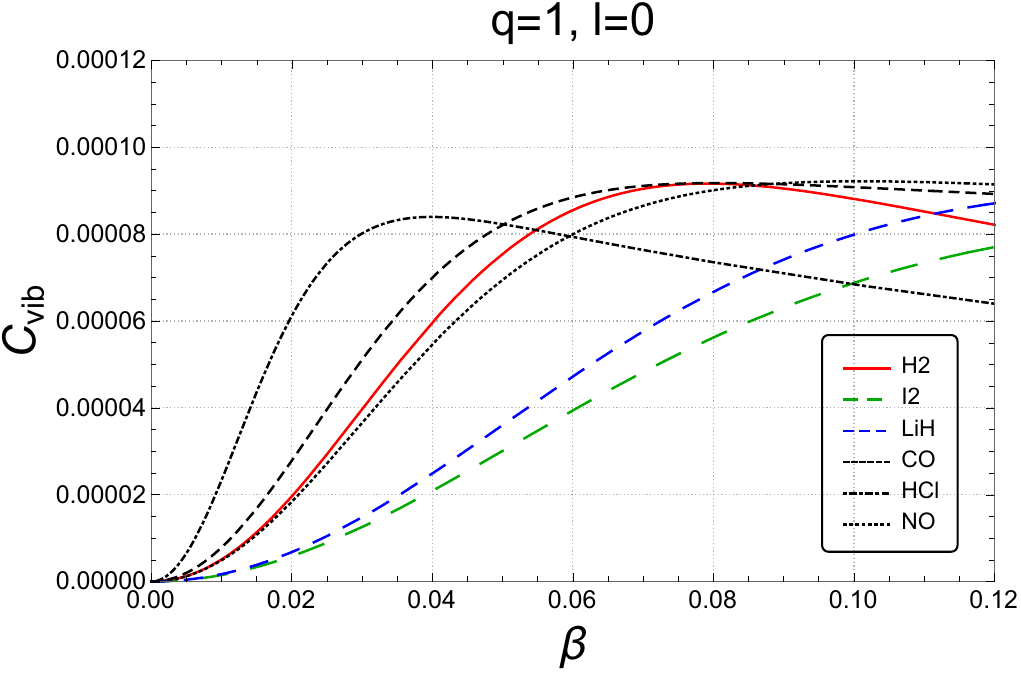}
	\includegraphics[width=0.33\textwidth]{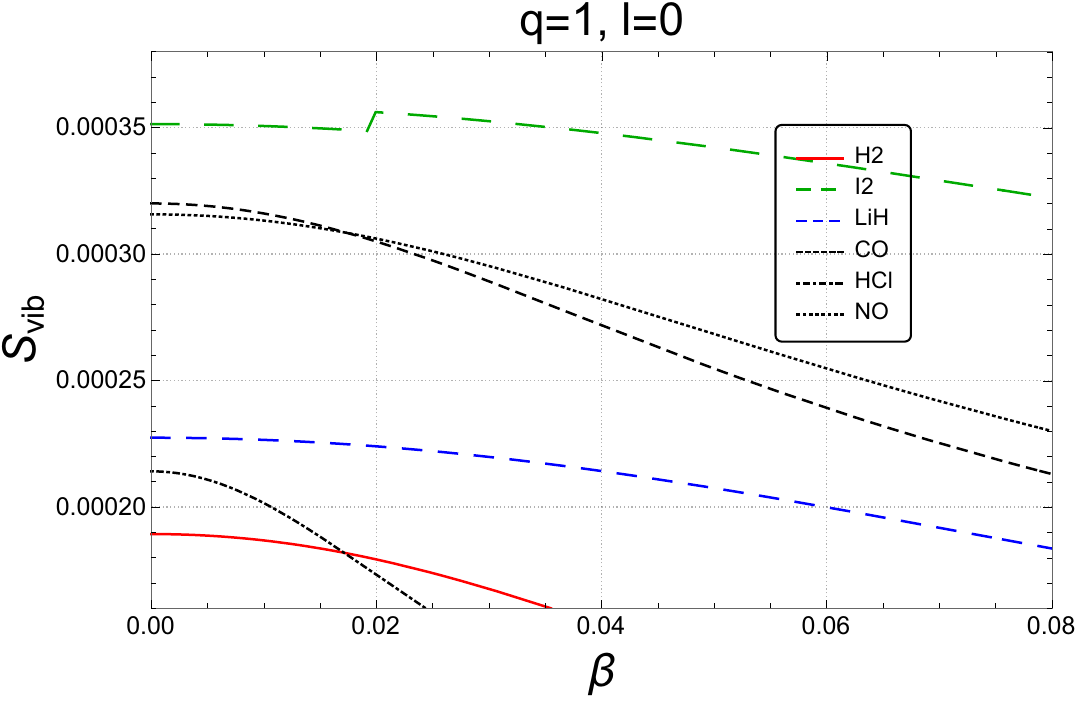}
	\caption{Variation of the energy spectrum and thermodynamic properties as a function of $q$ and $\beta$, respectively, for some diatomic molecules.}  \label{fig:z2}
\end{figure}

In Fig.~\ref{fig:z2}, we present the profiles of the energy spectrum and the vibrational thermodynamic properties, such as the partition function, free energy, mean energy, heat capacity, and entropy, as functions of the inverse temperature $\beta$ for the selected diatomic molecules $H_2$, $I_2$, $LiH$, $CO$, $HCl$, and $NO$. The general behavior of the thermodynamic functions is nearly the same for the chosen molecules; however, their variation with respect to $\beta$ differs. The values of these functions are very sensitive to changes in the potential parameters.

In the plots of the vibrational partition function $Z_{vib}$, we distinguish three types of behavior. The first is a slow variation with increasing $\beta$ (decreasing $T$), which corresponds to the molecules $H_2$ and $LiH$. These two molecules have similar values of the potential parameters $a/m$ and $b/m$ relative to the bond length $r_e$ and the screening parameter $\alpha$ (the parameters $a$ and $b$ are defined in the introduction). Note that the dipole moment is not taken into account in this model; it is zero for $H_2$ and very large for $LiH$. The second type of behavior is a moderate variation, corresponding to the molecules $CO$ and $NO$. Since the spectroscopic parameters of these two molecules are close, the values of $a/m$ and $b/m$ lie in the same range, with a larger difference between $a/m$ and $b/m$ than in the cases of $H_2$ and $LiH$. The small difference between $CO$ and $NO$ comes from the dissociation energy $D_e$, which is larger for $CO$, and the screening parameter $\alpha$, which is larger for $NO$. This is due to the type of bonding: $CO$ has a triple bond, whereas $NO$ has a double bond with a free radical. Again, note that $CO$ and $NO$ are polar molecules, and polarity is not taken into account. The last type of behavior is a rapid variation with respect to $\beta$. The corresponding molecules are $I_2$ and $HCl$. $I_2$ is much heavier than the other molecules and shows the smallest value of the parameter $b/m$, due to the small values of $D_e$ and $\alpha$ and the large value of $m$. On the other hand, $HCl$ shows the largest gap between the values of the parameters $a$ and $b$ for a non-heavy molecule. It also has a relatively large bond length $r_e$ but a deeper potential. The value of $r_e$ is related to the strong polarity of $HCl$. This suggests that the Yukawa term $e^{-\alpha c r/r}$ dominates for small $r$, which leads to this rapid variation.\\
The Helmholtz vibrational free energy $F_{vib}$ has the same qualitative behavior for all molecules, since it directly reflects the potential energy with a negative sign. Moreover, the vibrational energy $U_{vib}$ shows the same general behavior for all molecules; the difference lies in the numerical values. We observe that, for $LiH$, $H_2$, $NO$, and $CO$, the magnitude $\vert U_{vib} \vert$ increases as the potential depth $D_e$ increases. This is no longer the case for $I_2$ and $HCl$, since the first molecule is heavy and the second one exhibits a very rapid variation of $Z_{vib}$.\\
The vibrational specific heat $C_{vib}$ represents the contribution of vibrations to the total specific heat. For all the chosen molecules, the curves exhibit a maximum value that separates two regimes: a high-temperature regime (small $\beta$), in which the behavior decreases as a function of temperature, and a low-temperature regime (large $\beta$), in which the behavior increases as a function of temperature. The position of this maximum shifts from one molecule to another. The maximum is nearly the same for the molecules $H_2$, $NO$, and $CO$, although the spectroscopic parameters of $H_2$ are not close to those of $NO$ and $CO$. The same may be said for the molecules $I_2$ and $LiH$. 

The last set of graphs represents the vibrational entropy. Here, we see that a drop occurs at relatively high temperatures (small $\beta$) for the molecules $HCl$ and $H_2$. For the other molecules, this drop occurs at lower temperatures (larger $\beta$), which may be attributed to the stronger influence of the deformation parameter $q$ on the $HCl$ and $H_2$ molecules. This drop can be interpreted as the vanishing of $S_{vib}$, which represents the contribution to disorder arising from vibrational energy. Vibrations contribute to disorder at high temperature (low $\beta$). At low temperature, other phenomena, such as rotations, contribute to the entropy.

%---------------------------------------------------------------------------------------
\section{Conclusion}
%---------------------------------------------------------------------------------------

In this paper, we have investigated the path-integral treatment of a linear combination of Yukawa and four-parameter potentials used as a model for diatomic molecule interactions. As shown above, the Green’s function associated with this potential cannot be evaluated in a unified manner for arbitrary values of the deformation parameter. By introducing an appropriate approximation to treat the centrifugal term, we obtained the energy spectrum and the normalized wave functions of the bound states from the poles of the Green’s function and their residues. The partition function, derived from the energy spectrum, allowed us to evaluate several thermodynamic properties. To verify the accuracy of our results, we performed a numerical analysis of the energy levels as functions of the deformation and screening parameters for several diatomic molecules, namely $H_2$, $I_2$, $LiH$, $CO$, $HCl$, and $NO$, and we presented the corresponding plots of the various thermodynamic functions.

	%-------------------------------------------------------------------------------------------------%-------------------------------------------------------------------------------------------------
	
	%---------------------------------------------------------------------------------------
	
	\bibliographystyle{unsrt}  
	\bibliography{references}

@article{Malli,
	title={Molecular integrals involving hulthen-type functions (n=l STO) in relativistic quantum chemistry},
	author={Malli, Gulzari},
	journal={Chemical Physics Letters},
	volume={78},
	number={3},
	pages={578--580},
	year={1981},
	publisher={Elsevier}
}

@article{Myhrman,
	title={Exact eigenvalues, eigenfunctions, matrix elements and phase-shifts for a particle of angular momentum l in a particular screened potential},
	author={Myhrman, Ulla},
	journal={Journal of Mathematical Physics},
	volume={21},
	number={7},
	pages={1732--1739},
	year={1980},
	publisher={American Institute of Physics}
}

@article{Arai1,
	title={Exact solutions of multi-component nonlinear Schrodinger and Klein-Gordon equations in two-dimensional space-time},
	author={Arai, Asao},
	journal={Journal of Physics A: Mathematical and General},
	volume={34},
	number={20},
	pages={4281},
	year={2001},
	publisher={IOP Publishing}
}

@article{Hulthen,
	title={On the virtual state of the deuteron},
	author={Hulth{\'e}n, Lamek},
	journal={Physical Review},
	volume={61},
	number={9-10},
	pages={671},
	year={1942},
	publisher={APS}
}

@article{Saxon,
	title={Diffuse surface optical model for nucleon-nuclei scattering},
	author={Woods, Roger D and Saxon, David S},
	journal={Physical Review},
	volume={95},
	number={2},
	pages={577},
	year={1954},
	publisher={APS}
}

@article{Yukawa,
	title={On the interaction of elementary particles. I},
	author={Yukawa, Hideki},
	journal={Proceedings of the Physico-Mathematical Society of Japan. 3rd Series},
	volume={17},
	pages={48--57},
	year={1935},
	publisher={The Physical Society of Japan, The Mathematical Society of Japan}
}

@article{Kratzer,
	title={Die ultraroten rotationsspektren der halogenwasserstoffe},
	author={Kratzer, Adolf},
	journal={Zeitschrift f{\"u}r Physik},
	volume={3},
	pages={289--307},
	year={1920},
	publisher={Springer-Verlag}
}

@article{Hellman,
	title={A new approximation method in the problem of many electrons},
	author={Hellmann, Hans},
	journal={The Journal of Chemical Physics},
	volume={3},
	number={1},
	pages={61--61},
	year={1935},
	publisher={American Institute of Physics}
}

@article{Falaye1,
	title={Exact solution of Schr{\"o}dinger equation with q-deformed quantum potentials using Nikiforov—Uvarov method},
	author={Falaye, BJ and Oyewumi, KJ and Abbas, M},
	journal={Chinese Physics B},
	volume={22},
	number={11},
	pages={110301},
	year={2013},
	publisher={IOP Publishing}
}

@article{Ahmadov,
	title={Analytical solutions of the Schr{\"o}dinger equation for the Hulth{\'e}n potential within SUSY quantum mechanics},
	author={Ahmadov, HI and Jafarzade, Sh I and Qocayeva, MV},
	journal={International Journal of Modern Physics A},
	volume={30},
	number={32},
	pages={1550193},
	year={2015},
	publisher={World Scientific}
}

@article{Miraboutalebi,
	title={Solutions of N-dimensional Schr{\"o}dinger equation with Morse potential via Laplace transforms},
	author={Miraboutalebi, S and Rajaei, L},
	journal={Journal of Mathematical Chemistry},
	volume={52},
	pages={1119--1128},
	year={2014},
	publisher={Springer}
}

@article{Falaye2,
	title={Any l-state solutions of the Eckart potential via asymptotic iteration method},
	author={Falaye, Babatunde J},
	journal={Central European Journal of Physics},
	volume={10},
	pages={960--965},
	year={2012},
	publisher={Springer}
}

@article{Ibekwe,
	title={Mass spectrum of heavy quarkonium for screened Kratzer potential (SKP) using series expansion method},
	author={Ibekwe, EE and Okorie, US and Emah, JB and Inyang, EP and Ekong, SA},
	journal={The European Physical Journal Plus},
	volume={136},
	pages={1--11},
	year={2021},
	publisher={Springer}
}

@article{Inyang,
	title={Solutions of the Schr{\"o}dinger equation with Hulth{\'e}n-screened Kratzer potential: application to diatomic molecules},
	author={Inyang, Etido and Iwuji, PC and Ntibi, Joseph E and William, ES and Ibanga, EA},
	journal={East European Journal of Physics},
	number={2},
	pages={12--22},
	year={2022}
}

@article{Onyenegecha1,
	title={Approximate solutions of Schrodinger equation and expectation values of Inversely Quadratic Hellmann-Kratzer (IQHK) potential},
	author={Onyenegecha, CP and Okereke, CJ and Njoku, IJ and Madu, CA and Ndubuisi, RU and Nwajeri, UK},
	journal={The European Physical Journal Plus},
	volume={137},
	number={1},
	pages={147},
	year={2022},
	publisher={Springer}
}

@article{William,
	title={Arbitrary {\ensuremath{\ell}}-solutions of the Schr{\"o}dinger equation interacting with Hulth{\'e}n-Hellmann potential model},
	author={William, ES and Inyang, EP and Thompson, EA},
	journal={Revista mexicana de f{\'\i}sica},
	volume={66},
	number={6},
	pages={730--741},
	year={2020},
	publisher={Sociedad Mexicana de F{\'\i}sica}
}

@article{Ita,
	title={Approximate solution of the Schr{\"o}dinger equation with Manning-Rosen plus Hellmann potential and its thermodynamic properties using the proper quantization rule},
	author={Louis, Hitler and Ita, Benedict I and Nzeata, Nelson I},
	journal={The European Physical Journal Plus},
	volume={134},
	number={7},
	pages={315},
	year={2019},
	publisher={Springer Berlin Heidelberg}
}

@article{Onyenegecha2,
	title={Approximate solutions of Schr{\"o}dinger equation for the Hua plus modified Eckart potential with the centrifugal term},
	author={Onyenegecha, CP and Ukewuihe, UM and Opara, AI and Agbakwuru, CB and Okereke, CJ and Ugochukwu, NR and Okolie, SA and Njoku, IJ},
	journal={The European Physical Journal Plus},
	volume={135},
	number={7},
	pages={1--10},
	year={2020},
	publisher={Springer}
}

@article{Ukewuihe,
	title={Approximate solutions of Schrodinger equation in D Dimensions with the modified Mobius square plus Hulthen potential},
	author={Ukewuihe, UM and Onyenegecha, C Paul and Udensi, SC and Nwokocha, CO and Okereke, C Jennifer and Njoku, IJ and Iloanya, AC},
	journal={Mathematics and Computational Sciences},
	volume={2},
	number={2},
	pages={1--15},
	year={2021},
	publisher={Qom University of Technology}
}

@article{Onyenegecha3,
	title={Solutions of Schrodinger equation for the modified Mobius square plus Kratzer potential},
	author={Onyenegecha, CP and Onate, CA and Echendu, OK and Ibe, AA and Hassanabadi, H},
	journal={The European Physical Journal Plus},
	volume={135},
	number={3},
	pages={1--9},
	year={2020},
	publisher={Springer}
}

@article{Edet,
	title={Any l-state solutions of the Schr{\"o}dinger equation for q-deformed Hulthen plus generalized inverse quadratic Yukawa potential in arbitrary dimensions},
	author={Edet, CO and Okoi, PO},
	journal={Revista mexicana de f{\'\i}sica},
	volume={65},
	number={4},
	pages={333--344},
	year={2019},
	publisher={Sociedad Mexicana de F{\'\i}sica}
}

@book{Feynman,
	title={Quantum mechanics and path integrals},
	author={Feynman, Richard P and Hibbs, Albert R and Styer, Daniel F},
	year={2010},
	publisher={Courier Corporation}
}

@article{Sadoun1,
	title={Exact path integral treatment of a diatomic molecule potential},
	author={Benamira, F and Guechi, L and Mameri, S and Sadoun, MA},
	journal={Journal of mathematical physics},
	volume={48},
	number={3},
	year={2007},
	publisher={AIP Publishing}
}

@article{Sadoun2,
	title={Unified path integral treatment for generalized Hulth{\'e}n and Woods--Saxon potentials},
	author={Benamira, F and Guechi, L and Mameri, S and Sadoun, MA},
	journal={Annals of Physics},
	volume={322},
	number={9},
	pages={2179--2194},
	year={2007},
	publisher={Elsevier}
}

@article{Guechi1,
	title={Complete non-relativistic bound state solutions of the Tietz-Wei potential via the path integral approach},
	author={Khodja, A and Kadja, A and Benamira, F and Guechi, L},
	journal={The European Physical Journal Plus},
	volume={134},
	pages={1--12},
	year={2019},
	publisher={Springer}
}

@article{Diaf1,
	title={Unified treatment of the bound states of the Schi{\"o}berg and the Eckart potentials using Feynman path integral approach},
	author={Diaf, A},
	journal={Chinese Physics B},
	volume={24},
	number={2},
	pages={020302},
	year={2015},
	publisher={IOP Publishing}
}

@book{Grosche1,
	title={Handbook of Feynman path integrals},
	author={Grosche, Christian and Steiner, Frank and Steiner, Frank},
	volume={145},
	year={1998},
	publisher={Springer}
}

@book{Kleinert,
	title={Path integrals in quantum mechanics, statistics, polymer physics, and financial markets},
	author={Kleinert, Hagen},
	year={2009},
	publisher={World scientific}
}

@article{Arai2,
	title={Exactly solvable supersymmetric quantum mechanics},
	author={Arai, Asao},
	journal={Journal of Mathematical Analysis and Applications},
	volume={158},
	number={1},
	pages={63--79},
	year={1991},
	publisher={Elsevier}
}

@article{Diaf2,
	title={Feynman integral treatment of the Rosen--Morse potential with a centrifugal term approximation},
	author={Diaf, Ahmed and Hachama, Mohamed},
	journal={Canadian Journal of Physics},
	volume={91},
	number={12},
	pages={1081--1085},
	year={2013},
	publisher={NRC Research Press}
}

@article{Guechi2,
	title={Path integral solution for a deformed radial Rosen--Morse potential},
	author={Kadja, A and Benamira, F and Guechi, L},
	journal={Indian Journal of Physics},
	volume={91},
	pages={259--262},
	year={2017},
	publisher={Springer}
}

@article{Mustapic,
	title={Summing the spectral representations of P{\"o}schl--Teller and Rosen--Morse fixed-energy amplitudes},
	author={Kleinert, H and Mustapic, I},
	journal={Journal of mathematical physics},
	volume={33},
	number={2},
	pages={643--662},
	year={1992},
	publisher={American Institute of Physics}
}

@book{Gradshtein,
	title={Table of integrals, series, and products},
	author={I. S. Gradshtein and I. M. Ryzhik},
	year={2007},
	publisher={Academic Press, New York}
}

@article{Cai,
	title={Path-integral treatment of the Hulth{\'e}n potential},
	author={Cai, JM and Cai, PY and Inomata, A},
	journal={Physical Review A},
	volume={34},
	number={6},
	pages={4621},
	year={1986},
	publisher={APS}
}

@article{Boudjedaa,
	title={Path integral treatment for a screened potential},
	author={Boudjedaa, T and Chetouani, L and Guechi, L and Hammann, TF},
	journal={Journal of mathematical physics},
	volume={32},
	number={2},
	pages={441--446},
	year={1991},
	publisher={American Institute of Physics}
}

@article{Chetouani2,
	title={Exact path integral solution for a screened potential},
	author={Chetouani, L and Guechi, L and Lecheheb, A and Hammann, TF and Messouber, A},
	journal={Nuovo Cimento, B},
	volume={113},
	year={1998}
}

@article{Sadoun3,
	title={Path Integral Treatment of a Linear Combination of Deformed Diatomic Molecule Potentials},
	author={Sadoun, Mohamed Am{\'e}ziane and Touati, Abdellah},
	journal={International Journal of Theoretical Physics},
	volume={63},
	number={3},
	pages={67},
	year={2024},
	publisher={Springer}
}

@article{Sadoun4,
	title={Relativistic bound states solutions with a linear combination of Yukawa and deformed Hulth{\'e}n potentials by path integral approach},
	author={Sadoun, Mohamed Ameziane and Adnane, Hamza},
	journal={International Journal of Geometric Methods in Modern Physics},
	volume={21},
	number={1},
	pages={2450025--36},
	year={2024}
}

@article{Sadoun5,
	title={Path integral solutions for Klein-Gordon particle with position-dependent mass in deformed Hulth{\'e}n potential},
	author={Sadoun, MA},
	journal={Europhysics Letters},
	volume={142},
	number={3},
	pages={30001},
	year={2023},
	publisher={IOP Publishing}
}

@article{vibpartitionf2,
	title={Partition function of improved Tietz oscillators},
	author={Jia, Chun-Sheng and Wang, Chao-Wen and Zhang, Lie-Hui and Peng, Xiao-Long and Zeng, Ran and You, Xu-Tao},
	journal={Chemical Physics Letters},
	volume={676},
	pages={150--153},
	year={2017},
	publisher={Elsevier}
}

@article{vibpartitionf1,
	title={An accurate closed-form expression for the partition function of Morse oscillators},
	author={Strekalov, ML},
	journal={Chemical physics letters},
	volume={439},
	number={1-3},
	pages={209--212},
	year={2007},
	publisher={Elsevier}
}

@article{thermodynamic2020,
	title={Thermodynamic properties of improved deformed exponential-type potential (IDEP) for some diatomic molecules},
	author={Okorie, Uduakobong S and Ikot, Akpan N and Chukwuocha, Ephraim O and Rampho, GJ},
	journal={Results in Physics},
	volume={17},
	pages={103078},
	year={2020},
	publisher={Elsevier}
}

@article{thermodynamic2018,
	title={Thermodynamics properties of diatomic molecules with general molecular potential},
	author={Ikot, Akpan N and Chukwuocha, EO and Onyeaju, MC and Onate, CA and Ita, BI and Udoh, ME},
	journal={Pramana},
	volume={90},
	pages={1--9},
	year={2018},
	publisher={Springer}
}

@article{thermodynamic2017,
	title={Approximate bound-states solution of the Dirac equation with some thermodynamic properties for the deformed Hylleraas plus deformed Woods-Saxon potential},
	author={Onyeaju, MC and Ikot, AN and Onate, CA and Ebomwonyi, O and Udoh, ME and Idiodi, JOA},
	journal={The European Physical Journal Plus},
	volume={132},
	pages={1--18},
	year={2017},
	publisher={Springer}
}

@article{thermodynamic2019,
	title={Approximate solution of the Schr{\"o}dinger equation with Manning-Rosen plus Hellmann potential and its thermodynamic properties using the proper quantization rule},
	author={Louis, Hitler and Ita, Benedict I and Nzeata, Nelson I},
	journal={The European Physical Journal Plus},
	volume={134},
	number={7},
	pages={315},
	year={2019},
	publisher={Springer Berlin Heidelberg}
}

@article{thermodynamic2022,
	title={Analytical solutions of the N-dimensional Schr{\"o}dinger equation with modified screened Kratzer plus inversely quadratic Yukawa potential and thermodynamic properties of selected diatomic molecules},
	author={Inyang, Etido P and Ayedun, Funmilayo and Ibanga, Efiong A and Lawal, Kolawole M and Okon, Ituen B and William, Eddy S and Ekwevugbe, Omugbe and Onate, Clement A and Antia, Akaninyene D and Obisung, Effiong O},
	journal={Results in Physics},
	volume={43},
	pages={106075},
	year={2022},
	publisher={Elsevier}
}

@article{thermodynamic2018-2,
	title={A study of thermodynamic properties of quadratic exponential-type potential in D-dimensions},
	author={Okorie, Uduakobong Sunday and Ikot, Akpan N and Onyeaju, MC and Chukwuocha, EO},
	journal={Revista Mexicana de fisica},
	volume={64},
	number={6},
	pages={608--614},
	year={2018},
	publisher={Sociedad Mexicana de F{\'\i}sica}
}

@article{thermodynamic2018-3,
	title={Thermodynamic properties of the modified Yukawa potential},
	author={Okorie, US and Ibekwe, EE and Ikot, AN and Onyeaju, MC and Chukwuocha, EO},
	journal={Journal of the Korean physical society},
	volume={73},
	pages={1211--1218},
	year={2018},
	publisher={Springer}
}

@article{spectroscopic1,
	title={Energy spectra and the expectation values of diatomic molecules confined by the shifted Deng-Fan potential},
	author={Oluwadare, OJ and Oyewumi, KJ},
	journal={The European Physical Journal Plus},
	volume={133},
	number={10},
	pages={422},
	year={2018},
	publisher={Springer Berlin Heidelberg}
}

@article{thermodynamic2023,
	title={Arbitrary l-state solutions of the Klein--Gordon equation with the Eckart plus a class of Yukawa potential and its non-relativistic thermal properties},
	author={Demirci, Mehmet and Sever, Ramazan},
	journal={The European Physical Journal Plus},
	volume={138},
	number={5},
	pages={1--17},
	year={2023},
	publisher={Springer}
}

@article{bayrak2007,
	title={Bound state solutions of the Hulth{\'e}n potential by using the asymptotic iteration method},
	author={Bayrak, ORHAN and Boztosun, I},
	journal={Physica Scripta},
	volume={76},
	number={1},
	pages={92--96},
	year={2007}
}

@article{greene1,
	title={Variational wave functions for a screened Coulomb potential},
	author={Greene, RL and Aldrich, C},
	journal={Physical Review A},
	volume={14},
	number={6},
	pages={2363},
	year={1976},
	publisher={APS}
}

@article{adebimpe1,
	title={Eigensolutions, scattering phase shift and thermodynamic properties of Hulthen-Yukawa potential},
	author={Adebimpe, O and Onate, CA and Salawu, SO and Abolanriwa, A and Lukman, AF},
	journal={Results in Physics},
	volume={14},
	pages={102409},
	year={2019},
	publisher={Elsevier}
}

@article{strekalov1,
	title={An accurate closed-form expression for the partition function of Morse oscillators},
	author={Strekalov, ML},
	journal={Chemical physics letters},
	volume={439},
	number={1-3},
	pages={209--212},
	year={2007},
	publisher={Elsevier}
}

\end{document}